\documentclass[a4paper, 12pt]{article}
\usepackage{geometry}
\usepackage{amsthm,amsmath,latexsym,amssymb,amsfonts,amscd}
\usepackage{graphics,lscape,fancyhdr,array,stmaryrd,euscript,wrapfig}
\usepackage{empheq,wrapfig}
\usepackage{verbatim,slashed}
\numberwithin{equation}{section}
\usepackage{hyperref,setspace}
\usepackage[skip=12pt, indent=18pt]{parskip}
\usepackage{tikz-cd}
\usepackage{mathrsfs}
\usepackage[numbers,sort&compress]{natbib}
\usepackage[nottoc]{tocbibind}

\definecolor{dark-red}{rgb}{0.4,0.15,0.15}
\definecolor{dark-blue}{rgb}{0.15,0.15,0.4}
\definecolor{medium-blue}{rgb}{0,0,0.5}
\definecolor{blue}{rgb}{0,0,1}
\hypersetup{colorlinks, linkcolor={blue}, citecolor={blue}, urlcolor={blue}}

\newcommand{\pl}{\partial}

\newcommand{\besubeqs}{\begin{subequations}}
\newcommand{\esubeqs}{\end{subequations}}

\newcommand{\grad}{{\partial}}
\newcommand{\divg}{{\partial\cdot}}

\newcommand{\eom}{\mathcal{E}}

\newcommand{\bbl}{[\![}
\newcommand{\bbr}{]\!]}

\newcommand{\Bigbbl}{\Big[\!\!\Big[}
\newcommand{\Bigbbr}{\Big]\!\!\Big]}

\usepackage{mathtools}

\newcommand{\ceil}[1]{\left\lceil #1 \right\rceil}
\newcommand{\floor}[1]{\left\lfloor #1 \right\rfloor}

\usepackage{times}

\begin{document}

%%%%%%%%%%%%%%%%%%%%%%%%%%%%%%%%%%%%%%%%%%%%%%%%%%%%%%%%%%%%%

\pagenumbering{gobble}
\hfill
\vskip 0.01\textheight

%%%%%%%%%%%%%%%%%%%%%%%%%%%%%%%%%%%%%%%%%%%%%%%%%%%%%%%%%%%%%

\begin{center}

{\Large\bfseries 
All actions for free massive higher-spin fields}

\vspace{0.4cm}

\vskip 0.03\textheight
\renewcommand{\thefootnote}{\fnsymbol{footnote}}
William \textsc{Delplanque}
\renewcommand{\thefootnote}{\arabic{footnote}}
\setcounter{footnote}{0}
\vskip 0.03\textheight

{\em Service de Physique de l'Univers, Champs et Gravitation, \\ Universit\'e de Mons, 20 place du Parc, 7000 Mons, 
Belgium}

\end{center}

%%%%%%%%%%%%%%%%%%%%%%%%%%%%%%%%%%%%%%%%%%%%%%%%%%%%%%%%%%%%%

\vskip 0.02\textheight

\begin{abstract}
Massive higher-spin states/fields appear in the effective description of various systems from hadrons and nuclei to black holes, whenever the point-particle approximation is justified, as well as in the bottom-up approaches to the quantum gravity problem. In four dimensions the actions for massive higher-spin fields utilize either the Singh-Hagen/Zinoviev set of auxiliary fields or a single chiral field, which is an $sl(2,\mathbb{C})$ spin-tensor of type $(2s,0)$, generalizing the Chalmers-Siegel approach. We show that these two actions are on the opposite ends of a discrete family of actions where the physical field is a spin-tensor of type $(s+k,s-k)$. The $(2s-1,1)$- and $(2s-2,2)$-cases generalize the Proca and the Fierz-Pauli actions, respectively, to all spins. A similar family of second-order actions exists for fermionic higher-spin fields.
\end{abstract}

%%%%%%%%%%%%%%%%%%%%%%%%%%%%%%%%%%%%%%%%%%%%%%%%%%%%%%%%%%%%%
\newpage
% \tableofcontents
% \newpage
%%%%%%%%%%%%%%%%%%%%%%%%%%%%%%%%%%%%%%%%%%%%%%%%%%%%%%%%%%%%%
\pagenumbering{arabic}
\setcounter{page}{1}

%%%%%%%%%%%%%%%%%%%%%%%%%%%%%%%%%%%%%%%%%%%%%%%%%%%%%%%%%%%%%
%%%%%%%%%%%%%%%%%%%%%%%%%%%%%%%%%%%%%%%%%%%%%%%%%%%%%%%%%%%%%
\section{Introduction}
\label{sec:Intro}
%%%%%%%%%%%%%%%%%%%%%%%%%%%%%%%%%%%%%%%%%%%%%%%%%%%%%%%%%%%%%
Whenever the size of an object is small compared to other relevant length scales one can apply the point-particle approximation and employ the poweful QFT/amplitude tools to describe the system within the effective theory paradigm, which can be followed by taking the classical limit of the amplitudes. As a recent example, the QFT/amplitude methods have been applied to the problem of dynamics of black holes, see \emph{e.g.} \cite{Buonanno:2022pgc,Guevara:2018wpp,Chung:2018kqs,Cangemi:2022bew,Cangemi:2023bpe,Skvortsov:2023jbn,Cangemi:2023ysz}. The same techniques and ideas can be applied to hadrons, nuclei, \emph{etc}. One can, of course, also consider the actual elementary particles. Particles in flat spacetime have to fall into the classification of unitary irreducible representations of the spacetime symmetry group \cite{Wigner:1939cj}, the Poincare group, and have two characteristics: mass and spin.\footnote{See \emph{e.g.} \cite{Bekaert:2006py,Basile:2016aen,Bekaert:2017khg} for reviews and generalizations of the classification.} Black holes as well as all macroscopic rotating objects correspond, from this point of view, to massive particles with very large quantum spin. The study of hadrons, as massive particles with arbitrary spin, resulted in the discovery of string theory as the primary candidate for the theory of quantum gravity. More generally, it seems that higher spin states may be necessary for the quantization of gravity.\footnote{The spectrum of string theory consists of infinitely many massive higher-spin states. Massless higher-spin states are tricky to introduce consistent interactions, but some toy models have been developed, \emph{e.g.} chiral higher-spin gravity \cite{Metsaev:1991mt,Metsaev:1991nb,Ponomarev:2016lrm,Skvortsov:2018jea,Skvortsov:2020wtf,Ponomarev:2017nrr,Krasnov:2021nsq,Sharapov:2022wpz}, see \cite{Bekaert:2022poo} for a review of the higher-spin gravity approach.} All of this motivates further study of massive higher-spin fields in the bottom-up approach.

The degrees of freedom of massive particles are classified by irreducible representations of the Wigner's little algebra, which is $so(d-1)$ in $d$ dimensions \cite{Wigner:1939cj}. One can loosely refer to such representations as spin, but in $d>4$ it is a sequence of (half)-integer numbers rather than just a single number. There is a great ambiguity in how to choose Lorentz-covariant field variables suitable for the description (equations of motion and action) of a given massive particle. In principle, any Lorentz (spin)-tensor whose decomposition under the little group contains a given spin may be chosen. The equations of motion are to project out the unwanted degrees of freedom and are easy to write. A suitable action is much more difficult to construct since the equations of motion may not admit an action with the same field content. Massive higher-spin fields as described by a symmetric and traceless rank-$s$ tensor (the symmetric approach) is such an example. Indeed, the equations of motion
\begin{align}
\big(\square-m^2\big)\Phi_{\mu_1...\mu_s} &= 0 \,, &
 \partial^{\nu}\Phi_{\nu\mu_2...\mu_{s}} &= 0\,,
\end{align}
appear non-Lagrangian since there are more equations than fields. The problem is not present for $s=0$, can be avoided for $s=1$ through the Proca action \cite{Proca_1936}, but requires auxiliary fields for $s\geq2$ and was first addressed for $s=2$ by Fierz and Pauli \cite{Fierz:1939ix} and later, for all spins, by Singh and Hagen \cite{Singh:1974qz,Singh:1974rc}. The Stueckelberg approach was extended to higher spins in $AdS_d$ by Zinoviev in \cite{Zinoviev:2001dt}, see also \cite{Zinovev:1983kh,Klishevich:1997pd} for the flat space. It requires more fields, but compensates for that by having a clear guiding principle -- gauge invariance. A closely related approach was proposed by Pashnev \cite{Pashnev:1989gm} and operates in terms of a quartet of unconstrained tensor fields, see \cite{Fegebank:2024yft} for the relation between the two. Other closely related ideas, all of which are based on symmetric tensors, include \cite{Pashnev:1989gm,Buchbinder:2005ua,Bekaert:2003uc, Francia:2010qp,Buchbinder:2007ix,Buchbinder:2008ss,Kaparulin:2012px,Kazinski:2005eb,Metsaev:2012uy,Skvortsov:2023jbn}. In particular, a geometric approach without auxiliary fields can be found in \cite{Francia:2007ee, Francia:2008ac}. Non-symmetric tensors were introduced in \cite{Curtright_1980413} through duality to describe massive spin-two fields and a conjectured extension to higher spins was put forward; see \cite{González_2008, Hell_2022, Abakumova:2023wve} for the recent related works. Alternatively, the light-cone approach to massive higher-spin fields has been developed in \cite{Metsaev:2005ar,Metsaev:2007rn,Metsaev:2022yvb}.

An additional option is available in $d=4$: one can choose a chiral field \cite{Chalmers:1997ui,Chalmers:2001cy,Ochirov:2022nqz}, which is a spin-tensor of $so(1,3)\sim sl(2,\mathbb{C})$ of type $(2s,0)$, \emph{i.e.} it is a rank-$2s$ symmetric tensor $\Phi_{A_1...A_{2s}}$, $A_i=1,2$. The key advantage is that the problematic transversality constraint is absent and the only equation of motion, the Klein-Gordon one, is, of course, Lagrangian. However, the parity is not easy to control. Nevertheless, it is with the chiral approach that the main progress has been made in the recent study of the black-hole scattering problem via QFT/amplitude methods, see \cite{Cangemi:2023bpe}. The symmetric approach stands only slightly behind, see \cite{Cangemi:2023ysz,Cangemi:2022bew,Skvortsov:2023jbn}.

Let us confine ourselves to $d=4$ from now on. The physical field in the symmetric approach is a type-$(s,s)$ spin-tensor $\Phi_{A_1...A_s,A'_1... A'_s}$, which is symmetric in both groups of indices. In principle, any type-$(s+k,s-k)$ spin-tensor with $k=0,...,s$ can play the role of a physical field. The chiral and the symmetric options stay on the opposite ends $k=s$ and $k=0$, respectively, with the latter requiring a number of auxiliary fields growing with $s$.

In the present paper, we bridge this gap and show that there are simple actions with the physical field of type $(s+k,s-k)$ for any $k$. The actions require auxiliary fields, whose number is $(s-k-1)$ for $k<s$. While $k=s$ is the chiral description, the cases $k=s-1$ and $k=s-2$ are also somewhat special: in the former the physical field is $\Phi_{A_1...A_{2s-1},A'}$, it does not require any auxiliary fields and it generalizes the Proca case to all spins; in the latter the physical field is $\Phi_{A_1...A_{2s-2},A'_1A'_2}$, one auxiliary field of type $(2s-4,0)$ is needed, which extends the Fierz-Pauli Lagrangian to all spins.

The outline of the paper is as follows. In Section \ref{sec:Fields} we review the Proca, Fierz-Pauli and Singh-Hagen ideas in more detail, which paves the way to the main result. The action is constructed in Section \ref{sec:Action}, while some technicalities are left to Appendix \ref{sec:app:proof}. In Section \ref{sec:Fermions} we show that fermionic higher-spin fields are also covered by the same (second-order) action. We conclude with some remarks in Section \ref{sec:Conclusion}. 

%%%%%%%%%%%%%%%%%%%%%%%%%%%%%%%%%%%%%%%%%%%%%%%%%%%%%%%%%%%%%
\section{Field content}
\label{sec:Fields}
%%%%%%%%%%%%%%%%%%%%%%%%%%%%%%%%%%%%%%%%%%%%%%%%%%%%%%%%%%%%%

\paragraph{Low spin examples.} Singh and Hagen \cite{Singh:1974qz,Singh:1974rc} constructed a Langrangian description of a free massive spin-$s$ field for any $s$. Let us focus on the bosonic case first, \cite{Singh:1974qz}. Singh and Hagen used the symmetric approach, where a massive (of mass $m$) spin-$s$ field is described by a traceless symmetric rank-$s$ tensor\footnote{Our convention is that all indices denoted by the same letter and are at the same level (up or down) are assumed to be symmetrized with the trace part subtracted. If a group of traceless symmetric indices $\mu_1...\mu_s$ belongs to the same tensor, we can simplify a bit more the notation by writing them as $\mu(s)$. For example:
\begin{align}
\Phi_{\mu_1\mu_2\mu_3} \equiv \Phi_{\mu\mu\mu} \equiv \Phi_{\mu(3)} \, . \notag
\end{align}
} $\Phi_{\mu(s)}$.
It satisfies the Klein-Gordon equation
\begin{align}
\big(\square - m^2\big)\Phi_{\mu(s)} = 0 \, ,
\label{eq:KG}
\end{align}
(where $\square := \partial_{\mu}\partial^{\mu}$) and propagates $2s+1$ degrees of freedom in $d=4$ provided that the transverse constraint is imposed
\begin{align}
\partial^{\mu}\Phi_{\mu(s)} = 0 \, .
\label{eq:transverse}
\end{align}
We need to construct a Lagrangian, whose equations of motion imply both \eqref{eq:KG} and \eqref{eq:transverse}. There is clearly a problem here because the number of equations exceeds the number of fields. For the case of a massive spin-$1$ field, the Proca action gives the equation of motion
\begin{align}
\square\Phi_{\mu} - \partial_{\mu}\partial^{\nu}\Phi_{\nu} - m^2\Phi_{\mu} = 0 \, ,
\end{align}
from which we deduce straightforwardly
\begin{align}
\partial^{\mu}\Phi_{\mu} = 0
\end{align}
by taking the divergence of the equation. Plugging this into the equation gives the expected Klein-Gordon equation. However, it becomes more difficult for $s>1$. Let us look at the case of a spin-$2$ field. The most general second order equation for a traceless rank-two tensor reads
\begin{align}
\big(\square-m^2\big)\Phi_{\mu(2)} + a\partial_{\mu}\partial^{\nu}\Phi_{\nu\mu} = 0 \, .
\end{align}
By taking its divergence, we get
\begin{align}
\big(1+\tfrac{a}{2}\big)\square\partial^{\alpha}\Phi_{\alpha\mu} + \tfrac{a}{2}\partial_{\mu}\partial^{\alpha}\partial^{\beta}\Phi_{\alpha\beta} - m^2\partial^{\nu}\Phi_{\nu\mu} = 0 \, ,
\end{align}
which leads to $\partial^{\nu}\Phi_{\nu\mu} = 0$ only if $a=-2$ and $\partial^{\alpha}\partial^{\beta}\Phi_{\alpha\beta} = 0$. This second condition cannot be obtained from the initial equation of motion. Therefore, it is impossible to construct a Lagrangian with only one traceless symmetric tensor $\Phi_{\mu\nu}$ for a spin-$2$ field, and for any spin higher than one. Fierz and Pauli \cite{Fierz:1939ix} proposed to introduce another field, an \emph{auxiliary field}, $\varphi$, of rank zero. Indeed, we need to have $\partial^{\alpha}\partial^{\beta}\Phi_{\alpha\beta} = 0$, which is a scalar constraint and, hence, the introduction of an auxiliary scalar field. The most general second-order equations of motion are therefore
\besubeqs
\begin{align}
\big(\square-m^2\big)\Phi_{\mu(2)} + a\partial_{\mu}\partial^{\nu}\Phi_{\nu\mu} + b\partial_{\mu}\partial_{\mu}\varphi &= 0 \, , 
\label{eq:SH_spin2_eq_main} \\
\big(\square-cm^2\big)\varphi + \partial^{\alpha}\partial^{\beta}\Phi_{\alpha\beta} &= 0 \, .
\label{eq:SH_spin2_eq_aux}
\end{align}
\esubeqs
Because the auxiliary field is not physical, one of the consequences of the equations of motion needs to be $\varphi=0$. Once we achieve that, the equation \eqref{eq:SH_spin2_eq_aux} implies $\partial^{\alpha}\partial^{\beta}\Phi_{\alpha\beta} = 0$. Therefore, we need to show that the auxiliary field vanishes. Let us remind that we need $a=-2$ for the divergence of the main equation \eqref{eq:SH_spin2_eq_main} (with $\varphi=0=\partial^{\alpha}\partial^{\beta}\Phi_{\alpha\beta}$) to give the transverse constraint $\partial^{\nu}\Phi_{\nu\mu}=0$. By taking the double-divergence of \eqref{eq:SH_spin2_eq_main} and then replacing $\partial^{\alpha}\partial^{\beta}\Phi_{\alpha\beta}$ via \eqref{eq:SH_spin2_eq_aux} we obtain
\begin{align}
(b-a-1)\square^2\varphi + m^2(ac+c+1)\square\varphi - cm^4\varphi = 0 \, ,
\end{align}
from which we deduce $\varphi = 0$ if we choose $b=-1$ and $c=1$. Thus, the Lagrangian description of a free massive spin-$2$ field can be achieved with just one auxiliary scalar field, which is usually packaged together into a traceful symmetric rank-two tensor. 
%%%%%%%%%%%%%%%%%%%%%%%%%%%%%%%%%%%%%%%%%%%%%%%%%%%%%%%%%%%%%

\paragraph{Singh-Hagen procedure.} As shown in the Singh-Hagen paper \cite{Singh:1974qz}, the spin-$3$ case needs the main (physical) field of rank three, $\Phi_{\mu\mu\mu}$, an auxiliary field of rank-one, $\Phi_{\mu}$, and another auxiliary field of rank zero, $\varphi$. Indeed, similarly to the spin-$2$ case, we try to get the transverse constraint $\pl^\nu\Phi_{\nu\mu\mu}=0$ by taking the divergence of the main equation of motion. However, it leads to the transverse constraint only if the double-divergence of the main field is zero,  $\pl^\nu\pl^\nu\Phi_{\nu\nu\mu}=0$. The double-divergence of the rank-$3$ field is a rank-$1$ object and, hence, we need to introduce a rank-$1$ auxiliary field $\Phi_{\mu}$ together with its equation of motion. Then, as for the spin-$2$, the vanishing of the auxiliary field implies $\pl^\nu\pl^\nu\Phi_{\nu\nu\mu}=0$. However, in this case, taking the double-divergence of the main equation of motion and using the other equation implies $\Phi_{\mu}=0$ only if $\pl^\nu \Phi_{\nu}=0$ and $\pl^\nu\pl^\nu\pl^\nu\Phi_{\nu\nu\nu}=0$. These two scalar conditions cannot be deduced from the equations of motion. Therefore, a scalar auxiliary field $\varphi$ is needed. It is then easy to understand the general pattern.

In order to construct a Lagrangian for a free massive spin-$s$ field, we need the following set of fields:
\begin{align}
\Phi_{\mu(s)}\ ,\, \Phi_{\mu(s-2)}\ ,\, \Phi_{\mu(s-3)}\ ,\cdots\ ,\, \Phi_{\mu}\ ,\, \Phi \, .
\end{align}
The first field is the main (physical) field, and the others are the auxiliary (non-physical) fields. We want their equations of motion to imply the following set of \textit{constraints}
\besubeqs
\begin{align}
&\bullet\; \big(\square-m^2\big)\Phi_{\mu(s)} = 0 \, \text{ : the Klein-Gordon equation,} \\
&\bullet\; \partial^{\mu}\Phi_{\mu(s)} = 0 \, \text{ : the transverse constraint,} \\
&\bullet\; \Phi_{\mu(s-2)} = 0 \, , \notag \\
&\bullet\; \Phi_{\mu(s-3)} = 0 \, , \notag \\
&\;\; \cdots \notag \\
&\bullet\; \Phi = 0 \, \text{ : the vanishing of all auxiliary (non physical) fields.}
\end{align}
\esubeqs
Remarkably, having all these constraints as consequences of the equations of motion fixes all the coefficients in the action. The procedure is as follows. First, we write the most general second-order Lagrangian (up to normalization of the fields) and we derive the equations of motion. Next, we form their \textit{consequences} by taking combinations of divergences and d'Alembertians of the equations of motion. We need one constraint of each rank from $0$ to $s-1$, and we proceed from the bottom to the top. If we manage to impose 
\begin{align}
\Phi_{\mu(j)}=0 \, ,
\end{align}
then, by plugging it into the equations of motion, we find
\begin{align}
\partial^{\mu_1} \cdots \partial^{\mu_k}\Phi_{\mu_1\cdots\mu_{k+j}} = 0 \, , \quad \forall k \in \bbl 0, s-j \bbr \ \backslash\ \{s-1-j\} \, .
\end{align}
Of course, the consequences of the rank-$0$, ... , rank-$(j-1)$ constraints are used to work with the rank-$j$ constraint. Finally, by going from rank-$0$ to rank-$(s-1)$, we obtain the 
vanishing of all the auxiliary fields (from rank-$0$ to rank-$(s-2)$ constraints) and the transverse constraint (the rank-$(s-1)$ constraint). A similar procedure works for fermions \cite{Singh:1974rc}.

%%%%%%%%%%%%%%%%%%%%%%%%%%%%%%%%%%%%%%%%%%%%%%%%%%%%%%%%%%%%%

\paragraph{Types of fields.} Singh and Hagen used what we call the \emph{symmetric approach} to describe massive fields, but there are many more Lorentz covariant fields that can, in principle, be used to describe/carry the same physical degrees of freedom. In order to explain this idea, let us first recall the two-component spinor language. As is well-known the Lorentz algebra $so(1,3)$ is isomorphic to $sl(2,\mathbb{C})$, which allows us to label the irreducible representations of the Lorentz algebra with a pair $(m,n)$ of non-negative integers. A field associated to the Lorentz representation $(m,n)$ is a spin-tensor\footnote{It is common to call ``tensors'' tensors with Lorentz indices, and ``spin-tensors'' tensors with $sl(2,\mathbb{C})$ indices, even they are both tensors.}
\begin{align}
\Phi_{A(m),A'(n)} \, ,
\end{align}
where $A,B,C,\cdots=1,2$ and $A',B',C',\cdots=1,2$ are the indices of the (anti-)fundamental representations of $sl(2,\mathbb{C})$, raised and lowered with the help of the anti-symmetric Levi-Civita symbol $\epsilon_{AB}$ (and $\epsilon_{A'B'}$). We choose the convention
\besubeqs
\begin{align}
v^A = \epsilon^{AB}v_{B} \, , \quad v_A = v^B\epsilon_{BA} \, &, \quad v^{A'} = \epsilon^{A'B'}v_{B'} \, , \quad v_{A'} = v^{B'}\epsilon_{B'A'} \, , \\
\quad \epsilon_A{}^B = \delta_A^B \, &, \quad \epsilon_{A'}{}^{B'} = \delta_{A'}^{B'} \, .
\end{align}
\esubeqs
The spin is the label of an irreducible representation of the Wigner's litter algebra $so(3)\sim su(2)$. However, the irreducible representation of the Lorentz algebra $(m,n)$ is not, in general, an irreducible representation of $su(2)\sim so(3)$, but can be decomposed as\footnote{We use the rank, $r$, of $su(2)$-tensors as the label of irreducible representations. }
\begin{align}
(m,n) = \bigoplus_{r=\tfrac{|m-n|}2}^{\tfrac{m+n}2}(r) \, .
\end{align}
Therefore, if we want to describe a massive spin-$s$ particle by a field of type $(m,n)$, the minimal option is to have $\ m+n = 2s\ $, but we still have, in general, too many components because this representation contains other spins, hence, the need of the transverse constraint to eliminate all the unwanted spins. Indeed, the divergence of $\Phi_{A(m),A'(n)}$ is a field of type $(m-1, n-1)$, which contains the same spins as $(m,n)$ except the spin-$s$. 

Singh and Hagen focused on the representation $(s,s)$, \emph{i.e.} the symmetric approach. The corresponding field is $\Phi_{A(s),A'(s)}$ or $\Phi_{\mu(s)}$ in the vectorial language. Another interesting choice is the representation $(2s, 0)$ (or, equivalently, $(0,2s)$), because it is also a spin-$s$ irreducible representation of $su(2)\sim so(3)$.\footnote{We do not discuss the reality conditions. In principle, any non-symmetric description has to be supplemented by certain reality conditions on the fields, but for some problems, \emph{e.g.} for the calculation of amplitudes, one can ignore them. } The corresponding field is $\Phi_{A(2s)}$ (or, equivalently, $\Phi_{A'(2s)}$) and has, therefore, the right number of degrees of freedom of a spin-$s$ particle, \emph{i.e.} $2s+1$. This approach we refer to as the \emph{chiral approach}, which was proposed in \cite{Ochirov:2022nqz} and for lower spins in \cite{Chalmers:1997ui,Chalmers:2001cy}. In the chiral approach, which also unifies bosons and fermions, the free Lagrangian is trivial because we do not need the transverse constraint
\begin{align}
\mathcal{L} = \tfrac{1}{2}\Phi^{A(2s)}\big(\square-m^2\big)\Phi_{A(2s)} \, .
\end{align}
The goal of this paper is to generalize the Singh-Hagen procedure to all approaches between the symmetric and chiral ones, where the physical field is of type $(m,n)$.

%%%%%%%%%%%%%%%%%%%%%%%%%%%%%%%%%%%%%%%%%%%%%%%%%%%%%%%%%%%%%

\begin{figure}[h!]
\centering
\begin{tikzpicture}[scale=0.75]
\draw[->,very thick] (0,0) - - (20,0) ;
\draw[->,very thick] (0,0) - - (0,11) ;
\draw[dashed] (8.5,9.5) - - (18,0) ;
\draw[dashed] (6.5,7.5) - - (14,0) ;
\draw[dashed] (5.5,6.5) - - (12,0) ;
\draw[dashed] (4.5,5.5) - - (10,0) ;
\draw[dashed] (3.5,4.5) - - (8,0) ;
\draw[dashed] (2.5,3.5) - - (6,0) ;
\draw[dashed] (1.5,2.5) - - (4,0) ;
\draw[dashed] (0.5,1.5) - - (2,0) ;
\draw[dashed] (-0.5,0.5) - - (0,0) ;
\node[above left] at (8.5,9.5) {$\Phi_s$} ;
\node[above left] at (6.5,7.5) {$\Phi_{s-2}$} ;
\node[above left] at (5.5,6.5) {$\Phi_{s-3}$} ;
\node[above left] at (3.7,5.3) {$\cdot $} ;
\node[above left] at (2.7,4.3) {$\cdot $} ;
\node[above left] at (1.7,3.3) {$\cdot $} ;
\node[above left] at (0.5,1.5) {$\Phi_1$} ;
\node[above left] at (-0.5,0.5) {$\Phi_0$} ;
\draw (0,0) - - (9,9) ;
\draw (2,0) - - (10,8) ;
\draw (4,0) - - (11,7) ;
\draw (6,0) - - (12,6) ;
\draw (8,0) - - (13,5) ;
\draw (10,0) - - (14,4) ;
\draw (12,0) - - (15,3) ;
\draw (14,0) - - (16,2) ;
\draw (16,0) - - (17,1) ;
\node[below] at (19.9,-0.2) {\Large $k$} ;
\node[left] at (-0.2,10.9) {\Large $\ell$} ;
\node[below left] at (0,0) {$0$} ;
\node at (9,9) {$\bullet$} ;
\node at (7,7) {$\bullet$} ;
\node at (6,6) {$\bullet$} ;
\node at (5,5) {$\bullet$} ;
\node at (4,4) {$\bullet$} ;
\node at (3,3) {$\bullet$} ;
\node at (2,2) {$\bullet$} ;
\node at (1,1) {$\bullet$} ;
\node at (0,0) {$\bullet$} ;
\node at (10,8) {$\bullet$} ;
\node at (8,6) {$\bullet$} ;
\node at (7,5) {$\bullet$} ;
\node at (6,4) {$\bullet$} ;
\node at (5,3) {$\bullet$} ;
\node at (4,2) {$\bullet$} ;
\node at (3,1) {$\bullet$} ;
\node at (2,0) {$\bullet$} ;
\node at (11,7) {$\bullet$} ;
\node at (9,5) {$\bullet$} ;
\node at (8,4) {$\bullet$} ;
\node at (7,3) {$\bullet$} ;
\node at (6,2) {$\bullet$} ;
\node at (5,1) {$\bullet$} ;
\node at (4,0) {$\bullet$} ;
\node at (12,6) {$\bullet$} ;
\node at (10,4) {$\bullet$} ;
\node at (9,3) {$\bullet$} ;
\node at (8,2) {$\bullet$} ;
\node at (7,1) {$\bullet$} ;
\node at (6,0) {$\bullet$} ;
\node at (13,5) {$\bullet$} ;
\node at (11,3) {$\bullet$} ;
\node at (10,2) {$\bullet$} ;
\node at (9,1) {$\bullet$} ;
\node at (8,0) {$\bullet$} ;
\node at (14,4) {$\bullet$} ;
\node at (12,2) {$\bullet$} ;
\node at (11,1) {$\bullet$} ;
\node at (10,0) {$\bullet$} ;
\node at (15,3) {$\bullet$} ;
\node at (13,1) {$\bullet$} ;
\node at (12,0) {$\bullet$} ;
\node at (16,2) {$\bullet$} ;
\node at (14,0) {$\bullet$} ;
\node at (17,1) {$\bullet$} ;
\node at (18,0) {$\bullet$} ;
\draw[dotted] (9,9) - - (9,0) ;
\draw[dotted] (9,9) - - (0,9) ;
\draw[dotted] (10,8) - - (10,0) ;
\draw[dotted] (10,8) - - (0,8) ;
\node[below] at (9,0) {$s$} ;
\node[left] at (0,9) {$s$} ;
\node[below] at (10,0) {$s\! +\! 1$} ;
\node[left] at (0,8) {$s\! -\! 1$} ;
\node[below] at (18,0) {$2s$} ;
\draw[->] (9.3,9.3) - - (9.8,9.8) ;
\node[above right] at (9.8,9.8) {\footnotesize Symmetric approach} ;
\draw[->] (18.3,0.3) - - (18.8,0.8) ;
\node[above] at (19.5,0.8) {\footnotesize Chiral approach} ;
\draw[->, gray] (17.3,1.3) - - (17.8,1.8) ;
\node[above] at (18,1.8) {\footnotesize \color{gray}{Proca}} ;
\draw[->, gray] (16.3,2.3) - - (16.8,2.8) ;
\node[above] at (17.5,2.8) {\footnotesize \color{gray}{Fierz-Pauli}} ;
\end{tikzpicture}
\caption{\textit{The diagram depicts the types of fields involved in various approaches. The type of a field $\Phi_{A(k),A'(\ell)}$ is $(k,\ell)$, which is represented by a point on this picture. Each oblique solid line links all the fields (depicted by bullets) necessary for a given approach. As indicated, the top line (the biggest set of auxiliary fields) corresponds to the symmetric approach, and the last one (where there is no line because there is only one field) corresponds to the chiral approach. Each dashed oblique line links all the fields of a same rank.}}
\label{fig:figure1}
\end{figure}
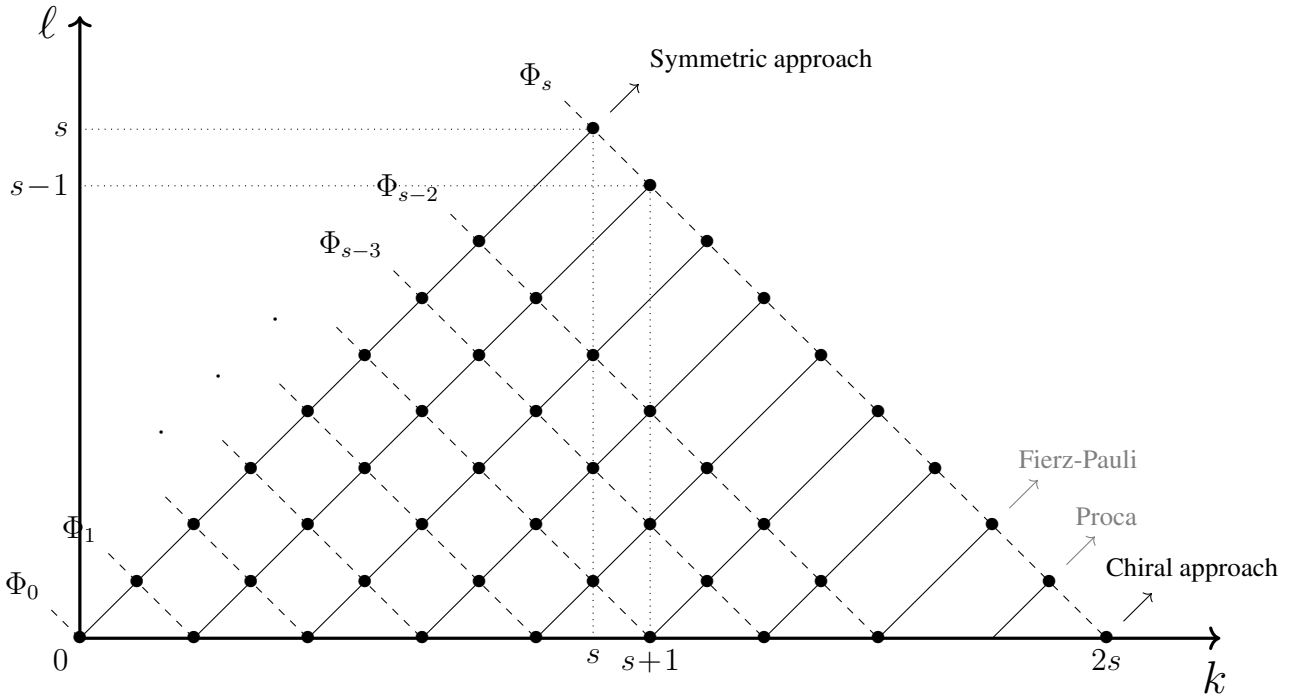

\paragraph{General pattern.} It is easy to guess the general pattern of auxiliary fields. Let us fix $m\geq n$ without loss of generality and choose the physical field to have type-$(m,n)$. We need to eliminate its divergence, \emph{i.e.} to get the transverse constraint, which is of type $(m-1,n-1)$. Following the Singh-Hagen idea we can see that this can be achieved as soon as the double-divergence vanishes, which has type-$(m-2,n-2)$. Therefore, we need to introduce an auxiliary field of same type. However, we need to also ensure that both the divergence of the auxiliary field and the triple-divergence of the physical one vanish, which calls for an auxiliary field of type $(m-3,n-3)$ and so on, see Figure \ref{fig:figure1} for a visualization of this procedure. The procedure needs $n-1$ auxiliary fields (or zero if $n=0$). The complete set of fields to have an action reads thus
\begin{align}
(m,n) \oplus (m-2,n-2)\oplus (m-3,n-3)\oplus (m-4,n-4) \oplus \cdots \oplus (m-n,0) \, .
\end{align}

For example, choosing the main field $\Phi_{A(s-1),A'}$ should be very similar to the spin-$1$ case in the symmetric approach, for any spin $s$. It does not require any auxiliary field. Choosing the field $\Phi_{A(s-2),A'(2)}$ should be similar to the spin-$2$ case in the symmetric approach because we will need one auxiliary field $\Phi_{A(s-2)}$ to get the transverse constraint, for any spin $s$.

In more detail, we write the most general ansatz for the second-order action. We want that the corresponding set of equations be equivalent to the Klein-Gordon equation for the main field, and $n$ constraints (with ranks from $(m-n,0)$ to $(m-1,n-1)$), which are the vanishing of all auxiliary fields and the transverse constraint for the main field. From the type-$(m-n+j,j)$ constraint we should deduce
\besubeqs
\begin{align}
\Phi_{A(m-n+j),A'(j)} &= 0 \, , \\
\partial^{A_1A'_1}\cdots\partial^{A_kA'_k}\Phi_{A(m-n+j+k),A'(j+k)} &= 0 \, , \quad \forall k \in \bbl 0, n-j \bbr \ \backslash\ \{n-1-j\}\, ,
\end{align}
\label{eq:result_constraint_general}
\esubeqs
where we used $sl(2,\mathbb{C})$ indices to write $\partial_{\mu} \sim \partial_{AA'}$. It is advantageous to use the results of the rank-$(j-1)$ constraint when working with the rank-$j$ one. As in the Singh-Hagen case, we expect this procedure to fix all coefficients in the action.

In order to simplify the formulas later on, let us introduce the rank as the half of the total number of indices, \emph{i.e.} $r=\tfrac{k+\ell}{2}$ for a field of type $(k,\ell)$. It is also convenient to define the \emph{chirality} $\lambda := \frac{m-n}{2}$, which is the same both for the physical field and for the auxiliary fields. Therefore, we can refer to a description based on the physical field of type $(m,n)$ as to chirality-$\lambda$ approach. 

For example, the chirality of the symmetric approach is equal to zero, and equal to the spin for the chiral approach. With these redefinitions, the set of fields that we need in the action depends on the spin $s$ and the chirality $\lambda$ and contains the fields 
\begin{align}
\Phi_s\ ,\,\, \Phi_{s-2}\ ,\,\, \Phi_{s-3}\ ,\,\, \cdots\ ,\,\, \Phi_{\lambda+1}\ ,\,\, \Phi_{\lambda} \, ,
\end{align}
where we indicated the rank of each field as a subscript. Starting from the physical field down to the last auxiliary one, we see that the ranks of the fields are the same for all approaches (all chiralities), but we keep only $s-\lambda-1$ auxiliary fields, see Figure \ref{fig:figure1}.
%%%%%%%%%%%%%%%%%%%%%%%%%%%%%%%%%%%%%%%%%%%%%%%%%%%%%%%%%%%%%

%%%%%%%%%%%%%%%%%%%%%%%%%%%%%%%%%%%%%%%%%%%%%%%%%%%%%%%%%%%%%

In this section, we justified the content of auxiliary fields for a description based on the type-$(m,n)$ physical field. In the next section, we construct the action and show how to fix the coefficients.

%%%%%%%%%%%%%%%%%%%%%%%%%%%%%%%%%%%%%%%%%%%%%%%%%%%%%%%%%%%%%
%%%%%%%%%%%%%%%%%%%%%%%%%%%%%%%%%%%%%%%%%%%%%%%%%%%%%%%%%%%%%
\section{Action}
\label{sec:Action}
%%%%%%%%%%%%%%%%%%%%%%%%%%%%%%%%%%%%%%%%%%%%%%%%%%%%%%%%%%%%%

\paragraph{Index-free notation and generating functions.} A convenient way to deal with a number of (spin)-tensors is to introduce generating functions, which hides all the indices and will also make all formulas more compact. In particular we will see that all equations and constraints will only depend on the rank of the fields, rather than on the chirality, which unifies all approaches to describes the same spin-$s$ degrees of freedom.  

Let us introduce auxiliary variables $y^A$ and $\bar{y}^{A'}$. A field $\Phi_{A(k),A'(\ell)}$ is replaced by a generating function $\Phi$ of the variables $y^A$ and $\bar{y}^{A'}$
\begin{align}
\Phi \equiv \Phi(y, \bar{y}) := \Phi_{A(k),A'(\ell)}\; y^A\cdots y^A\ \bar{y}^{A'}\cdots \bar{y}^{A'} \, .
\label{eq:def_Phi_generating}
\end{align}
Since the variables are commuting, the Taylor coefficients of generating functions are automatically symmetric. We can define the derivatives with respect to these variables:
\begin{align}
\partial_A \equiv \frac{\partial}{\partial y^A} \, , \qquad\qquad \bar{\partial}_{A'} \equiv \frac{\partial}{\partial\bar{y}^{A'}} \, .
\end{align}
We can extract the type of the field with the help of the operators
\begin{align}
N := y^A\partial_A \, , \qquad\qquad \bar{N} := \bar{y}^{A'}\bar{\partial}_{A'} \, ,
\end{align}
which count, respectively, the number of unprimed and primed indices of a field. Indeed it is easy to show that
\begin{align}
N\Phi = k\Phi \, , \qquad\qquad \bar{N}\Phi = \ell\Phi \, .
\end{align}
However, $N$ does not commute with $y^A$ and $\partial_A$
\begin{align}
[N,y^A]\Phi = y^A\Phi \, , \qquad\qquad [N,\partial_A]\Phi = -\partial_A\Phi \, ,
\label{eq:commut_N_y_d}
\end{align}
\emph{idem} with $\bar{N}$, $\bar{y}^{A'}$ and $\bar{\partial}_{A'}$. The action and equations of motion contain also derivatives (with respect to spacetime coordinates) of fields, so it is necessary to define these operators in the index-free notation. Let us define the gradient and the divergence respectively as
\begin{align}
\grad := \big(N\bar{N}\big)^{\tfrac{1}{2}}\ y^A\bar{y}^{A'}\partial_{AA'} \, , \qquad\qquad \divg := \big((N+1)(\bar{N}+1)\big)^{-\tfrac{1}{2}}\ \partial_A\bar{\partial}_{A'}\partial^{AA'} \, .
\label{eq:def_grad_div}
\end{align}
Due to \eqref{eq:commut_N_y_d}, $N$ and $\bar{N}$ do not commute with them
\besubeqs
\begin{align}
\big[ N,\grad\big] &= -\grad \, , \qquad\qquad \big[ N,\divg\big] = \divg \, , \\
\big[\bar{N},\grad\big] &= -\grad \, , \qquad\qquad \big[\bar{N},\divg\big] = \divg \, .
\end{align}
\label{eq:commut_N_grad_div}
\esubeqs
By using \eqref{eq:def_grad_div} and \eqref{eq:commut_N_grad_div}, it is straightforward to show that
\begin{align}
\big[\divg , \grad\big]\Phi = \Big(1+\tfrac{N+\bar{N}}{2}\Big)\square\Phi \, ,
\label{eq:commut_div_grad}
\end{align}
where $\square := \partial_{AA'}\partial^{AA'}$.\footnote{Note that our convention for the translation between spinorial and vectorial languages for tensors remains such that the scalars are identical: $v_{\mu}v^{\mu} \equiv v^2 \equiv v_{AA'}v^{AA'}$. Recall also that an anti-symmetric spin-tensor is proportional to its trace: $T_{[AB]} \equiv \tfrac{1}{2}T_C{}^C\epsilon_{AB}$. Therefore 
\begin{align}
\partial_{AA'}\partial_{B}{}^{A'} \equiv \tfrac{1}{2}\epsilon_{AB}\square \, , \qquad\qquad \partial_{AA'}\partial^{A}{}_{B'} \equiv \tfrac{1}{2}\epsilon_{A'B'}\square \, , \notag
\end{align}
which is anti-symmetric on $A\leftrightarrow B$ (or on $A'\leftrightarrow B'$) because partial derivatives commute.} It is easy to extend this result recursively to obtain
\begin{align}
\big[\divg^{\ell} , \grad\big]\Phi = \Big(\ell\big(1+\tfrac{N+\bar{N}}{2}\big) + \tfrac{\ell(\ell-1)}{2}\Big)\square\divg^{\ell-1}\Phi \, ,
\label{eq:commut_div^l_grad}
\end{align}
and then
\begin{align}
\big[\divg^{\ell} , \grad^2\big]\Phi = \Big(\ell\big(1+\tfrac{N+\bar{N}}{2}\big) &+ \tfrac{\ell(\ell-1)}{2}\Big)\Big((\ell-1)\big(1+\tfrac{N+\bar{N}}{2}\big) + \tfrac{(\ell-1)(\ell-2)}{2}\Big)\square^2\divg^{\ell-2}\Phi \notag \\
&+\Big(\ell\big(1+N+\bar{N}\big)+\ell(\ell-1)\Big)\square\grad\divg^{\ell-1}\Phi \, .
\label{eq:commut_div^l_grad^2}
\end{align}
These results will be useful later, and we observe that these relations do not depend on the chirality but only on the rank of the field (because they depend only on $N+\bar{N}$). 

Finally, let us introduce a notation for the scalar product between fields (\emph{i.e.} contraction of all their indices)
\begin{align}
\Phi\cdot\Psi := \Phi^{A(k),A'(\ell)}\Psi_{A(k),A'(\ell)} \, ,
\label{eq:def_scalar_product}
\end{align}
or, equivalently, in the language of generating functions
\begin{align}
\Phi\cdot\Psi \equiv \Phi(\partial,\bar{\partial})\big((N!\bar{N}!)^{-1}\Psi(y,\bar{y})\big) \, .
\end{align}
Let us remark that this scalar product is symmetric both for bosons and fermions. Indeed
\begin{align}
\Phi\cdot\Psi = \Phi^{A(k),A'(\ell)}\Psi_{A(k),A'(\ell)} &= (-1)^{\sigma}\Psi_{A(k),A'(\ell)}\Phi^{A(k),A'(\ell)} \notag \\
&= (-1)^{\sigma}(-1)^{k+\ell}\Psi^{A(k),A'(\ell)}\Phi_{A(k),A'(\ell)} = (-1)^{\sigma}(-1)^{k+\ell}\Psi\cdot\Phi \, ,
\label{eq:scalar_product_sym}
\end{align}
where $\sigma$ equals to zero for bosons and one for fermions, because bosonic fields are Grassmann-even and fermionic fields are Grassmann-odd. Then, let us recall that indices are contracted using the anti-symmetric Levi-Civita symbol, $T_A{}^A = -T^A{}_A$, hence, the factor $(-1)^{k+\ell}$ at second line. For boson fields, $k+\ell \equiv 2r$ is even and for fermions, $2r$ is odd. Therefore, in both cases $(-1)^{\sigma}(-1)^{k+\ell} = 1$.

Note that the gradient and divergence operators defined at \eqref{eq:def_grad_div} differ from the natural operators in the index notation by a factor 
\besubeqs
\begin{align}
(\grad\Phi)_{A(k+1),A'(\ell+1)} &= \sqrt{(k+1)(\ell+1)}\ \partial_{AA'}\Phi_{A(k),A'(\ell)} \, , \\
(\divg\Phi)_{A(k-1),A'(\ell-1)} &= \sqrt{k\ell}\ \partial^{AA'}\Phi_{A(k),A'(\ell)} \, .
\end{align}
\esubeqs
The conventions that we chose for the definitions \eqref{eq:def_grad_div} together with the scalar product \eqref{eq:def_scalar_product} imply
\begin{align}
\int\Phi\cdot(\divg\Psi)\;\mathrm{d}^4x = -\int\Psi\cdot(\grad\Phi)\;\mathrm{d}^4x\; +\; \text{boundary terms} \, .
\end{align}
It is obvious that any expression in terms of generating functions can be converted into the form with explicit indices modulo some factors, as explained above. 

%%%%%%%%%%%%%%%%%%%%%%%%%%%%%%%%%%%%%%%%%%%%%%%%%%%%%%%%%%%%%

\paragraph{Construction of the action.} Let us denote $\Phi_r$ the field of rank-$r$, \emph{i.e.} the field $\Phi_{A(k),A'(\ell)}$ where $k+\ell=2r$. The most general second-order equation $\mathcal{E}_r$ that has rank-$r$, but may involve fields of other ranks, has the form
\begin{align}
\mathcal{E}_r := a_r\square\Phi_r + b_r\grad\divg\Phi_r + c_r\grad^2\Phi_{r-2} + d_r\divg^2\Phi_{r+2} + e_rm\grad\Phi_{r-1} + f_rm\divg\Phi_{r+1} + g_rm^2\Phi_r = 0 \, .
\end{align}
If the set of equations of motion $\{\eom_r=0\}_{r\in\bbl\lambda,s\bbr\backslash\{s-1\}}$ comes from an action, we have
\begin{align}
d_r = c_{r+2} \, , \qquad f_r = -e_{r+1} \, ,
\label{eq:condition_on_coeff_if_action}
\end{align}
and the action reads
\begin{align}
S = \tfrac{1}{2}\!\!\int\!\Big(\Phi_s\cdot\big(&a_s\square\Phi_s + b_s\grad\divg\Phi_s + 2c_s\grad^2\Phi_{s-2} + g_sm^2\Phi_s\big) \label{eq:action_general} \\
&+ \sum_{r = \lambda}^{s-2}\Phi_r\cdot\big(a_r\square\Phi_r + b_r\grad\divg\Phi_r + 2c_r\grad^2\Phi_{r-2} + 2e_rm\grad\Phi_{r-1} + g_rm^2\Phi_r\big)\Big)\mathrm{d}^4x \, . \notag
\end{align}
Not all of these coefficients are physically relevant because we can rescale some of them via rescaling of fields. With the help of this freedom let us normalize the fields as
\besubeqs
\begin{align}
&\bullet\; a_s = 1 \;\; \text{: It normalizes } \Phi_s \text{ up to a sign.} \\
&\bullet\; c_s = 1 \;\; \text{: It links the normalization of } \Phi_{s-2} \text{ to } \Phi_s. \\
&\bullet\; e_r = 1 \;\; \forall r \ \in \bbl \lambda + 1, s-2 \bbr \;\; \text{: It links } \Phi_{s-3}\text{, } \Phi_{s-4}\text{, ... , }\Phi_{\lambda} \text{ to } \Phi_{s-2}. \\
&\bullet\; g_s = -1 \;\; \text{: It makes the parameter } m \text{ be the mass of the main (physical) field } \Phi_s.
\end{align}
\label{eq:normalization}
\esubeqs
With this choice we have used all the freedom in rescaling fields up to an overall sign, the unfixed sign of $\Phi_s$, which does not even show up in the action. 

The set of equations of motion $\{\eom_r=0\}_{r\in\bbl\lambda,s\bbr\backslash\{s-1\}}$, which are obtained from the action, needs to be equivalent to the following set
\begin{align}
\Big\{\big(\square-m^2\big)\Phi_s = 0\; ,\quad \divg\Phi_s = 0\;\Big\}\cup\Big\{\Phi_r = 0\Big\}_{r\in\bbl\lambda,s-2\bbr} \, .
\end{align}
The first element of this set is the Klein-Gordon equation of motion for the main (physical) field, and the others are called constraints. The first constraint is the transverse constraint for the main field to get the right number of propagating degrees of freedom, and the others are the vanishing of the auxiliary (non-physical) fields. Note that if we apply the constraints to the first equation of motion of the set $\{\eom_r=0\}_{r\in\bbl\lambda,s\bbr\backslash\{s-1\}}$, we obtain the Klein-Gordon equation of motion for the main field. Therefore, the only thing we have to do now is to derive the constraints from the original set of equations of motion.

In order to do that, we take for each rank $j\in\bbl\lambda, s-1\bbr$ a linear combination of these equations and their derivatives (consequences) that have rank-$j$. As explained in the previous section, the idea is to apply the consequences starting from the smallest rank to the highest one, and to use the rank-$(j-1)$ consequences when we express the rank-$j$ ones, see equations \eqref{eq:result_constraint_general}. Therefore, only divergences of the equations of motion are needed.\footnote{Indeed, for the rank-$j$ consequences, the gradient of an expression needs an expression of rank-$(j-1)$, which is $\divg^{\ell}\eom_{j-1+\ell}$, for any $\ell$ from zero to $s-j$. This expression contains divergences of the fields: $\divg^{i}\Phi_{j-1+i}$ with $i\in\{\ell-2, \ell-1, \ell, \ell+1, \ell +2\}$, which have already been shown to vanish by the rank-$(j-1)$ consequences, see eqs. \eqref{eq:result_constraint_general}.} As explained in the previous section, the rank-$j$ consequences need to yield $\Phi_j=0$ first, after which we deduce the full set \eqref{eq:result_constraint_general}. The general form of these consequences is (there is no sum over the index $j\in\bbl\lambda, s-1\bbr$)\footnote{For $j=s-1$, the consequence is $\divg\eom_s = 0 \Leftrightarrow \divg\Phi_s = 0$, \emph{i.e.} the sought for constraint.}
\begin{align}
\sum_{\ell}\sum_{k}A^j_{\ell,k} \square^k\divg^{\ell}\eom_{j+\ell} = 0 \quad \Leftrightarrow \quad \Phi_j = 0 \, ,
\label{eq:constraint_generic_expr}
\end{align}
where $A^j_{\ell,k}$ are the coefficients of the right linear combination used to extract $\Phi_j = 0$ from the set of $\square^k\divg^{\ell}\eom_{j+\ell}$. Let us develop $\eom_{j+\ell}$ in terms of the fields
\begin{align}
\square^k\divg^{\ell}\eom_{j+\ell} =\ &\alpha_{\ell}^j \square^{k+1}\divg^{\ell}\Phi_{j+\ell} + \beta_{\ell}^j \square^{k+2}\divg^{\ell-2}\Phi_{j+\ell-2} + \eta_{\ell}^j \square^{k}\divg^{\ell+2}\Phi_{j+\ell+2} \notag \\
&+ \varepsilon_{\ell}^j \square^{k+1}\divg^{\ell-1}\Phi_{j+\ell-1} + \zeta_{\ell}^j \square^{k}\divg^{\ell+1}\Phi_{j+\ell+1} + \gamma_{\ell}^j \square^{k}\divg^{\ell}\Phi_{j+\ell} \, ,
\label{eq:constraint_general_term}
\end{align}
where we used the properties \eqref{eq:commut_div^l_grad} and \eqref{eq:commut_div^l_grad^2}, and where
\besubeqs
\begin{align}
\alpha_{\ell}^j &:= a_{j+\ell} + \tfrac{\ell}{2}(2j + \ell + 1)b_{j+l} \, , \\
\beta_{\ell}^j &:= \tfrac{\ell(\ell-1)}{4}(2j+\ell+1)(2j+\ell)c_{j+\ell} \, , \\
\eta_{\ell}^j &:= d_{j+\ell} \, , \\
\varepsilon_{\ell}^j &:= \tfrac{\ell}{2}(2j+\ell+1)m\ e_{j+\ell} \, , \\
\zeta_{\ell}^j &:= m\ f_{j+\ell} \, , \\
\gamma_{\ell}^j &:= m^2g_{j+\ell} \, ,
\end{align}
\label{eq:def_greek}
\esubeqs
and where\footnote{Let us recall that the symbol $\floor{x}$ means the floor of $x$, and $\ceil{x}$ means the ceiling of $x$.}
\begin{align}
j \in \bbl \lambda , s-1 \bbr \; , \quad \ell \in [\![0,s-j]\!] \ \backslash\ \{s-1-j\} \; , \quad k \in \Bigbbl 0, \floor{\tfrac{s-j-\ell}{2}} \Bigbbr \, .
\end{align}
Let us explain the range for the values of $k$, \emph{i.e.} the maximal order of the $\square$ that needs to be used. The linear combination \eqref{eq:constraint_generic_expr} of derivatives of the equations giving the constraint needs to cancel all the terms with derivatives of fields, because it has to imply the vanishing of the auxiliary field of the corresponding rank. However, \eqref{eq:constraint_general_term} shows that it produces the term $\square^{k+2}\divg^{\ell-2}\Phi_{j+\ell-2}$. Therefore if we begin with $k=0$ for the highest $\ell$ (\emph{i.e.} $\ell = s-j$), we will not be able to cancel this term if we keep $k=0$ for $l=s-2-j$. With $k=1$ for $l=s-2-j$, the term $\square^{k+2}\divg^{\ell-2}\Phi_{j+\ell-2}$ could be canceled by the term $\square^{k+1}\divg^{\ell-1}\Phi_{j+\ell-1}$ for the next $\ell$, $\ell = s-3-j$, but the term $\square^{k+1}\divg^{\ell-1}\Phi_{j+\ell-1}$ for this $\ell$ cannot be canceled. We need therefore to allow $k=2$ for $l=s-j-4$, \emph{etc}. It is easy then to guess the pattern. Let us call $K^j_{\ell}$ the maximal value of $k$ for a given $j$ and a given $\ell$. We have that $K^j_{s-j} = 0$, $K^j_{s-j-2} = 1$, $K^j_{s-j-3} = 1$, $K^j_{s-j-4} = 2$, $K^j_{s-j-5} = 2$, $K^j_{s-j-6} = 3$, \emph{etc}. Therefore, we have in general: $K^j_{\ell} = \floor{\tfrac{s-j-\ell}{2}}$. Note that nothing forbids $k$ to take higher values, but we will restrict ourselves to the minimal values of $k$ that are sufficient, which will be clear from the proof.

Let us rewrite the relation \eqref{eq:constraint_general_term} in a matrix notation, $\vec\eom= \mathbb{A}_j \vec{\Phi}$, where vector $\vec\eom$ contains all the required $\square^k\divg^{\ell}\eom_{j+\ell}$; $\vec{\Phi}$ contains all $\square^{\Tilde{k}}\divg^{\Tilde{\ell}}\Phi_{j+\Tilde{\ell}}$ that contribute to $\vec\eom$; the matrix $\mathbb{A}_j$ is purely numerical with entries from \eqref{eq:def_greek}. The key of the procedure is to perform a row reduction of $\mathbb{A}_j$ as to get the constraints, which will also fix all the free coefficients in the action. Let us consider some examples, which we arrange starting from the highest value of $j$. Note that we need to analyze the consequences for all $j \in \bbl \lambda , s-1 \bbr$, \emph{i.e.} the matrices $\mathbb{A}_j$, but each $\mathbb{A}_j$ does not, in fact, depend on $\lambda$ and the size of $\mathbb{A}_j$ increases as $j$ decreases.   

\paragraph*{Constraint matrix $\boldsymbol{\mathbb{A}_{s-1}}$ :}
It is the simplest constraint. Only $k=0$ and $\ell=1$ is used, giving only the term $\divg\eom_s$. By using the matrix notation, \eqref{eq:constraint_general_term} reads
\begin{align}
\begin{pmatrix}
\divg\eom_s
\end{pmatrix}
=
\begin{pmatrix}
\alpha_1^{s-1} & \gamma_1^{s-1}
\end{pmatrix}
\begin{pmatrix}
\square\divg\Phi_s \\
\divg\Phi_s
\end{pmatrix} \, .
\end{align}
The rank-$(s-1)$ constraint needs to imply the transverse constraint for the main field ($\divg\Phi_s$=0). Therefore, we need to impose
\begin{align}
\alpha_1^{s-1} = 0 \quad \Leftrightarrow \quad b_s = -\frac{1}{s} \, ,
\label{eq:b_s}
\end{align}
because $\vec\eom=0$ on-shell.

\paragraph*{Constraint matrix $\boldsymbol{\mathbb{A}_{s-2}}$ :}
This constraint involves two values of $\ell=0,2$. Let us split the matrix into blocks separated by the values of $\ell$ in order to clarify the structure. For $\ell=0$, $k=0,1$. The set of relations \eqref{eq:constraint_general_term} now reads\footnote{Let us split $\vec\eom$ into blocks from $\ell = s-j$ to $\ell=0$ from top to bottom, and inside each block from $k=K^j_{\ell}$ to $k=0$ from the top to the bottom. The vector $\vec \Phi$ is organized in the same way.}
\begin{align}
\begin{pmatrix}
\divg^2\eom_s \\ \hline
\square\eom_{s-2} \\
\eom_{s-2}
\end{pmatrix}
=
\left(\begin{array}{cc|ccc}
\alpha_2^{s-2} & \gamma_2^{s-2} & \beta_2^{s-2} & 0 & 0 \\ \hline
\eta_0^{s-2} & 0 & \alpha_0^{s-2} & \gamma_0^{s-2} & 0 \\
0 & \eta_0^{s-2} & 0 & \alpha_0^{s-2} & \gamma_0^{s-2}
\end{array}\right)
\begin{pmatrix}
\square\divg^2\Phi_s \\
\divg^2\Phi_s \\ \hline
\square^2\Phi_{s-2} \\
\square\Phi_{s-2} \\
\Phi_{s-2}
\end{pmatrix} \, .
\end{align}
In order to have the constraint $\Phi_{s-2}=0$ we need to take a linear combination of the rows of $\mathbb{A}_{s-2}$, \emph{c.f.} \eqref{eq:constraint_generic_expr}, and to set all its components except the last one (corresponding to $\Phi_{s-2}$ in $\vec\Phi$) to zero. Taking into account the solution \eqref{eq:b_s} for $\mathbb{A}_{s-1}$, we find
\besubeqs
\begin{align}
a_{s-2} &= -s(2s-1) \, , \\
g_{s-2} &= \tfrac{s^2(2s-1)}{s-1} \, , \\
A^{s-2}_{2,0} &= \tfrac{1}{m^2} \, , \\
A^{s-2}_{0,1} &= \tfrac{1}{m^2}\tfrac{s-1}{s} \, , \\
A^{s-2}_{0,0} &= 1 \, .
\end{align}
\label{eq:solution_s-2}
\esubeqs
The constraint $\divg^2\Phi_s=0$ follows now. Indeed, once $\Phi_{s-2}=0$ we can forget the last block-column of the matrix.

\paragraph*{Constraint matrix $\boldsymbol{\mathbb{A}_{s-3}}$ :}
There are three possible values for $\ell$ now, making three times three blocks in the matrix:
\begin{align}
\begin{pmatrix}
\divg^3\eom_s \\ \hline
\square\divg\eom_{s-2} \\
\divg\eom_{s-2} \\ \hline
\square\eom_{s-3} \\
\eom_{s-3}
\end{pmatrix}
=
\left(\begin{array}{cc|ccc|ccc}
\alpha_3^{s-3} & \gamma_3^{s-3} & \beta_3^{s-3} & 0 & 0 & 0 & 0 & 0 \\ \hline
\eta_1^{s-3} & 0 & \alpha_1^{s-3} & \gamma_1^{s-3} & 0 & \varepsilon_1^{s-3} & 0 & 0 \\
0 & \eta_1^{s-3} & 0 & \alpha_1^{s-3} & \gamma_1^{s-3} & 0 & \varepsilon_1^{s-3} & 0 \\ \hline
0 & 0 & 0 & \zeta_0^{s-3} & 0 & \alpha_0^{s-3} & \gamma_0^{s-3} & 0 \\
0 & 0 & 0 & 0 & \zeta_0^{s-3} & 0 & \alpha_0^{s-3} & \gamma_0^{s-3}
\end{array}\right)
\begin{pmatrix}
\square\divg^3\Phi_s \\
\divg^3\Phi_s \\ \hline
\square^2\divg\Phi_{s-2} \\
\square\divg\Phi_{s-2} \\
\divg\Phi_{s-2} \\ \hline
\square^2\Phi_{s-3} \\
\square\Phi_{s-3} \\
\Phi_{s-3}
\end{pmatrix} \, .
\end{align}
First, repeating the same procedure as for $\Phi_{s-2}=0$ constraint in the previous example we can arrive at $\Phi_{s-3}=0$. Taking into account the results from $\mathbb{A}_{s-1}$ and $\mathbb{A}_{s-2}$, it is straightforward to find
\besubeqs
\begin{align}
b_{s-2} &= -s \, , \\
a_{s-3} &= -\tfrac{(s-1)(2s-3)}{s^3} \, , \\
g_{s-3} &= \tfrac{3(s-1)^3}{s^3(s-2)} \, , \\
A^{s-3}_{3,0} &= \tfrac{1}{m^3}\tfrac{s-1}{s^2 (2s-1)} \, , \\
A^{s-3}_{1,1} &= \tfrac{1}{m^3}\tfrac{(s-1)(2s-3)}{s^3 (2s-1)} \, , \\
A^{s-3}_{1,0} &= \tfrac{1}{m}\tfrac{s-1}{s^2(2s-1)} \, , \\
A^{s-3}_{0,1} &= \tfrac{1}{m^2}\tfrac{s-2}{2s-1} \, , \\
A^{s-3}_{0,0} &= 1 \, .
\end{align}
\label{eq:solution_s-3}
\esubeqs
Second, once $\Phi_{s-3}=0$ we can forget the last block-column and we deduce directly that $\divg\Phi_{s-2} = 0$. Lastly, we can forget the second block-column to obtain $\divg^3\Phi_{s}=0$.

\paragraph*{Constraint matrix $\boldsymbol{\mathbb{A}_{s-4}}$ :}
Similarly to the examples above, the matrix form of \eqref{eq:constraint_general_term} reads
{\scriptsize
\begin{align}
\begin{pmatrix}
\divg^4\eom_s \\ \hline
\square\divg^2\eom_{s-2} \\
\divg^2\eom_{s-2} \\ \hline
\square\divg\eom_{s-3} \\
\divg\eom_{s-3} \\ \hline
\square^2\eom_{s-4} \\
\square\eom_{s-4} \\
\eom_{s-4}
\end{pmatrix}
=
\left(\begin{array}{cc|ccc|ccc|cccc}
\alpha_4^{s-4} & \gamma_4^{s-4} & \beta_4^{s-4} & 0 & 0 & 0 & 0 & 0 & 0 & 0 & 0 & 0 \\ \hline
\eta_2^{s-4} & 0 & \alpha_2^{s-4} & \gamma_2^{s-4} & 0 & \varepsilon_2^{s-4} & 0 & 0 & \beta_2^{s-4} & 0 & 0 & 0 \\
0 & \eta_2^{s-4} & 0 & \alpha_2^{s-4} & \gamma_2^{s-4} & 0 & \varepsilon_2^{s-4} & 0 & 0 & \beta_2^{s-4} & 0 & 0 \\ \hline
0 & 0 & 0 & \zeta_1^{s-4} & 0 & \alpha_1^{s-4} & \gamma_1^{s-4} & 0 & 0 & \varepsilon_1^{s-4} & 0 & 0 \\
0 & 0 & 0 & 0 & \zeta_1^{s-4} & 0 & \alpha _1^{s-4}& \gamma_1^{s-4} & 0 & 0 & \varepsilon_1^{s-4} & 0 \\ \hline
0 & 0 & \eta_0^{s-4} & 0 & 0 & \zeta_0^{s-4} & 0 & 0 & \alpha_0^{s-4} & \gamma_0^{s-4} & 0 & 0 \\
0 & 0 & 0 & \eta_0^{s-4} & 0 & 0 & \zeta_0^{s-4} & 0 & 0 & \alpha_0^{s-4} & \gamma_0^{s-4} & 0 \\
0 & 0 & 0 & 0 & \eta_0^{s-4} & 0 & 0 & \zeta_0^{s-4} & 0 & 0 & \alpha_0^{s-4} & \gamma_0^{s-4}
\end{array}\right)
\begin{pmatrix}
\square\divg^4\Phi_s \\
\divg^4\Phi_s \\ \hline
\square^2\divg^2\Phi_{s-2} \\
\square\divg^2\Phi_{s-2} \\
\divg^2\Phi_{s-2} \\ \hline
\square^2\divg\Phi_{s-3} \\
\square\divg\Phi_{s-3} \\
\divg\Phi_{s-3} \\ \hline
\square^3\Phi_{s-4} \\
\square^2\Phi_{s-4} \\
\square\Phi_{s-4} \\
\Phi_{s-4}
\end{pmatrix} . \notag
\end{align}
}
This example illustrates the difficulty of the general case. Indeed, once $\Phi_{s-4}=0$ is achieved, we cannot deduce directly $\divg\Phi_{s-3}=0$ because of $\eta_0^{s-4}$ in the last line. We have to make another linear combination of the remaining rows, which is true for all $j\leq s-4$. In fact, we will set $\eta_0^j = 0$ for $j\leq s-4$, which simplifies the procedure and gives a solution. It implies that we need to have $d_j = 0$ for $j\leq s-4$, or, equivalently, (because it comes from an action) $c_j = 0$ for $j\leq s-2$. Therefore, due to $c_j=d_j = 0$, the consequence  \eqref{eq:constraint_general_term} for $j\leq s-2$ simplifies to
\begin{align}
\square^k\divg^{\ell}\eom_{j+\ell} = \alpha_{\ell}^j \square^{k+1}\divg^{\ell}\Phi_{j+\ell} + \varepsilon_{\ell}^j \square^{k+1}\divg^{\ell-1}\Phi_{j+\ell-1} + \zeta_{\ell}^j \square^{k}\divg^{\ell+1}\Phi_{j+\ell+1} + \gamma_{\ell}^j \square^{k}\divg^{\ell}\Phi_{j+\ell} \, .
\end{align}
As a result, we do not need to increase the maximal value of $k$ with $\ell$. Indeed, for $\ell=s-j$ we have only $k=0$, for $\ell=s-2-j$ the maximal value of $k$ is one, but after that no more $\square$'s are generated, and, hence, the maximal value of $k$ is one.

\paragraph*{Constraint matrix $\boldsymbol{\mathbb{A}_{j}}$ :}
The form of the matrix $\mathbb{A}_{j}$ is, schematically,
\begin{align}
\left(\begin{array}{cc|ccc|ccc|ccc|ccc|ccc|ccc|ccc}
\alpha & \gamma & \beta & 0 & 0 &  &  &  &  &  &  &  &  &  &  &  &  &  &  &  \\ \hline
\eta & 0 & \alpha & \gamma & 0 & \varepsilon & 0 & 0 &  &  &  &  &  &  &  &  &  &  &  &  \\
0 & \eta & 0 & \alpha & \gamma & 0 & \varepsilon & 0 &  &  &  &  &  &  &  &  &  &  &  &  \\ \hline
 &  & 0 & \zeta & 0 & \alpha & \gamma & 0 & \varepsilon & 0 & 0 &  &  &  &  &  &  &  &  &  \\
 &  & 0 & 0 & \zeta & 0 & \alpha & \gamma & 0 & \varepsilon & 0 &  &  &  &  &  &  &  &  &  \\ \hline
 &  &  &  &  & 0 & \zeta & 0 & \alpha & \gamma & 0 & \varepsilon & 0 & 0 &  &  &  &  &  &  \\
 &  &  &  &  & 0 & 0 & \zeta & 0 & \alpha & \gamma & 0 & \varepsilon & 0 &  &  &  &  &  &  \\ \hline
 &  &  &  &  &  &  &  &  0 & \zeta & 0 & \alpha & \gamma & 0 & \varepsilon & 0 & 0  &  &  &  &  &  & \\
 &  &  &  &  &  &  &  & 0 & 0 & \zeta & 0 & \alpha & \gamma & 0 & \varepsilon & 0  &  &  &  &  &  & \\ \hline
 &  &  &  &  &  &  &  &  &  &  &  0 & \zeta & 0 & \alpha & \gamma & 0 & \varepsilon & 0 & 0  &  &  & \\
 &  &  &  &  &  &  &  &  &  &  & 0 & 0 & \zeta & 0 & \alpha & \gamma & 0 & \varepsilon & 0 &  &  & \\ \hline
 &  &  &  &  &  &  &  &  &  &  &  &  &  &  0 & \zeta & 0 & \alpha & \gamma & 0 & \varepsilon & 0 & 0 \\
 &  &  &  &  &  &  &  &  &  &  &  &  &  & 0 & 0 & \zeta & 0 & \alpha & \gamma & 0 & \varepsilon & 0 \\ \hline
 &  &  &  &  &  &  &  &  &  &  &  &  &  & & & & 0 & \zeta & 0 & \alpha & \gamma & 0 \\
 &  &  &  &  &  &  &  &  &  &  &  &  &  & & & & 0 & 0 & \zeta & 0 & \alpha & \gamma
\end{array}\right) \, ,
\label{eq:matrix_generic}
\end{align}
where we omitted indices, and the empty blocks contain only zeros. Each block-line is related to a value of $\ell$, from $\ell = s-j$ to $\ell = 0$ from the top to the bottom (skipping $\ell = s-j-1$). Inside each block, the lines are related to values of $k$, from $k=1$ to $k=0$ from the top to the bottom. 

As a summary: for a given $\lambda$, we write the constraints from the bottom to the top, \emph{i.e.} from rank-$\lambda$ to rank-$(s-1)$. Therefore, when we write the rank-$j$ constraint we already know from the previous ones that
\begin{align}
\divg^{\ell}\Phi_{j-1+\ell} = 0 \;\; \forall \ell \in \bbl 0, s-j \bbr \qquad \text{and} \qquad \Phi_r = 0 \;\; \forall r \in \bbl \lambda, j-1 \bbr \, ,
\end{align}
and from this rank-$j$ constraint we want to deduce 
\begin{align}
\divg^{\ell}\Phi_{j+\ell} = 0 \;\; \forall \ell \in \bbl 0, s-j \bbr \, .
\end{align}
The matrix form helps us to see what needs to be done. Firstly, we need to find a linear combination of its rows which gives a line containing zeros except for the last element\footnote{It implies that $g_r\neq0$ for all $r$.}, which implies $\Phi_j=0$. Then, we need to deduce $\divg\Phi_{j+1} = 0$, $\divg^2\Phi_{j+2} = 0$, \emph{etc}. As long as we derive new constraints we can forget the last block-columns since they are now multiplied by zeros. After dropping these last block-columns we have directly the last line composed of only zeros except for the last element (which is non-zero because $f_j = -e_{j+1}$ is normalized to one), giving immediately $\divg\Phi_{j+1} = 0$, \emph{etc}. In other words, once $\Phi_j=0$ is achieved, thanks to its stair-like form the row reduction of the matrix becomes trivial.

With the help of \eqref{eq:normalization}, \eqref{eq:b_s} and $c_j = 0$ for $j\leq s-2$, the action \eqref{eq:action_general}  becomes
\begin{align}
S = \tfrac{1}{2}\!\!\int\!\Big(\Phi_s\cdot\big((\square-m^2)&\Phi_s - \tfrac{1}{s}\grad\divg\Phi_s + 2\grad^2\Phi_{s-2}\big) \notag \\
&+ \sum_{r = \lambda}^{s-2}\Phi_r\cdot\big(a_r\square\Phi_r + b_r\grad\divg\Phi_r + 2m\grad\Phi_{r-1} + g_rm^2\Phi_r\big)\Big)\mathrm{d}^4x \, .
\label{eq:action_general_valued}
\end{align}
Firstly, we observe an elegant structure for all fields now: each field interacts only with the nearest neighbors. In particular, $\Phi_s$ interacts only with $\Phi_{s-2}$ (because $\Phi_{s-1}$ does not exist), and then $\Phi_{s-2}$ interacts only with $\Phi_{s-3}$ (because $c_{s-2}=0$), which interacts only with $\Phi_{s-4}$ (because $c_{s-3}=0$), \emph{etc}.

\noindent Secondly, we see that there are only three coefficients (for each rank) that need to be determined by the procedure: $a_r$, $b_r$, and $g_r$, which gives (The proof of this result is in Appendix \ref{sec:app:proof}.), for $r \in \bbl \lambda, s-3 \bbr$,
\besubeqs
\begin{align}
b_{s-2} &= -s \, , \\
b_{r} &= \frac{2(r+2)}{(s-r-2)(s+r+3)(r+3)}\frac{1}{b_{r+1}} \, , \label{eq:solution_b_r_b_r+1} \\
a_{r} &= (2r + 3)b_{r} \;\; \text{, also true for } r = s-2 \, , \label{eq:solution_a_r} \\
g_{r} &= -\frac{(r+2)(s-r)(s+r+1)}{2(r+1)}b_{r} \;\; \text{, also true for } r = s-2 \, .
\label{eq:solution_g_r}
\end{align}
\label{eq:solution}
\esubeqs
This is a closed system of recursive relations between all the coefficients in the action. It is possible to get a closed form for $b_r$
\begin{align}
b_{r} &= -(r+2)\prod_{i=r/2}^{\tfrac{s-4}{2}}\frac{(2i+4)^2(s-2i-3)(s+2i+4)}{(2i+3)^2(s-2i-2)(s+2i+3)} \quad \text{ for }s-r\text{ even only} \, . \label{eq:solution_b_r}
\end{align}

%%%%%%%%%%%%%%%%%%%%%%%%%%%%%%%%%%%%%%%%%%%%%%%%%%%%%%%%%%%%%
%%%%%%%%%%%%%%%%%%%%%%%%%%%%%%%%%%%%%%%%%%%%%%%%%%%%%%%%%%%%%
\section{Fermionic massive higher-spin fields}
\label{sec:Fermions}
%%%%%%%%%%%%%%%%%%%%%%%%%%%%%%%%%%%%%%%%%%%%%%%%%%%%%%%%%%%%%
In this paper, we considered the case of bosons, as mentioned at the beginning of Section \ref{sec:Fields}. This paper proposes a generalization of the Singh-Hagen procedure \cite{Singh:1974qz} for free massive bosonic fields. It may be interesting to consider free massive fermionic fields. The first idea would be to generalize the Singh-Hagen procedure for fermions, \cite{Singh:1974rc}, which uses the symmetric approach. In \cite{Singh:1974rc} the set of fields consists of $\{\Psi_r\}_{r\in\bbl 1/2, s\bbr}$ and their conjugates $\{\bar{\Psi}_r\}_{r\in\bbl 1/2, s\bbr}$\footnote{If $\Psi_r$ is a field of type $(m,n)$ with $m+n=2r$, $\bar{\Psi}_r$ is of type $(n,m)$.} and the action is first-order in derivatives. Such an action helps to control parity. However, in the non-symmetric approaches the parity is not manifest. Moreover, the first-order action for fermions cannot be written in terms of some set of fields and their conjugates. Lastly, let us mention that the Singh-Hagen actions for fermions and bosons are rather different. The difference between bosons and fermions is almost absent in the second-order formulation. 

Indeed, the general procedure we developed in this paper remains correct for fermions too! The index-free notation does not discriminate bosons and fermions. In particular, the scalar product is symmetric both for boson and fermion fields, see \eqref{eq:scalar_product_sym}.\footnote{It implies in particular that \eqref{eq:condition_on_coeff_if_action} remains true for fermions.} Therefore, all manipulations with the action and consequences remain valid. In particular, it is not necessary to assume that $s$ is an integer, it could be also half-integer. Indeed, the rank of the constraints ranges $j \in \bbl \lambda , s-1 \bbr$, but the important number is
\vspace{-5mm}
\begin{align}
s-j \in \bbl s-\lambda , 1 \bbr \, ,
\end{align}
which is an integer both for bosons and fermions. Consequently, the procedure we developed works both for bosons and fermions. As a result, we constructed a family of actions ranging from the symmetric to the chiral approaches, with the free parameters $s$ (the spin) and $\lambda$ (the chirality) which can take any integer or half-integer value (with $\lambda \leq s$), with $s-\lambda$ integer. The symmetric approach for fermions is included in this general procedure in the second-order form.

%%%%%%%%%%%%%%%%%%%%%%%%%%%%%%%%%%%%%%%%%%%%%%%%%%%%%%%%%%%%%
\section{Conclusion}
\label{sec:Conclusion}
%%%%%%%%%%%%%%%%%%%%%%%%%%%%%%%%%%%%%%%%%%%%%%%%%%%%%%%%%%%%%
In the present paper we have bridged the gap between the symmetric and chiral approaches by showing that there exists a family of Lagrangians where the physical field is a spin-tensor of type $(s+k,s-k)$ for any $k$. The symmetric and chiral cases correspond to $k=0$ and $k=s$, respectively. It is interesting that the Proca and the Fierz-Pauli Lagrangians have a higher-spin generalization, in the sense of featuring the same number of auxiliary fields, which is $0$ and $1$, respectively.\footnote{A generalization of the Proca description to higher spins was proposed by K. Krasnov and E. Skvortsov (unpublished).} They correspond to $k=s-1$ and $k=s-2$, respectively. It is also interesting that this family of Lagrangians has $s$, the spin, as an unrestricted positive parameter. In particular, it can be either integer or half-integer, unifying bosons and fermions with the same second-order actions, and extending this property of the chiral approach to all the other approaches. One might also be interested in \cite{Koutrolikos_2021}, which presents a supersymmetry generalization of the Singh-Hagen procedure.

In the recent \cite{Delplanque:2024enh} it was shown for spins up to $s=2$ that the symmetric and chiral approaches are related via a certain invertible change of variables (field redefinitions). The procedure requires the more steps the higher the spin. For example, for $s=2$ one needs two steps, the first one being a relation between the symmetric approach, based on the type-$(2,2)$ physical field, and the Proca-like, based on the type-$(3,1)$ field. The last step is the relation between the $(3,1)$ and the chiral, $(4,0)$, descriptions. Therefore, it seems that establishing the equivalence between the symmetric and chiral approaches requires all the intermediate descriptions constructed in the present paper. In particular, it implies that there is an invertible change of variables that relates the $(m,n)$-description to the $(m+1,n-1)$-description. Establishing the equivalence of these approaches has important applications. Indeed, while constructing consistent interactions (in the sense of preserving the number of physical degrees of freedom) is easy in the chiral approach, imposing the parity symmetry is difficult. The situation in the symmetric approach is the opposite: parity is easy, while constructing consistent interactions is hard, see \emph{e.g.} \cite{Zinoviev:2006im,Zinoviev:2008ck,Zinoviev:2009hu,Zinoviev:2010cr,Buchbinder:2012iz}. 

Therefore, we expect that the present paper provides a useful link connecting all Lagrangian descriptions of massive higher-spin particles. The next step would be to study the transfer of various interactions between the approaches, which should help to impose parity in the chiral approach and to introduce consistent interactions in the symmetric one, see \emph{e.g.} \cite{Delplanque:2024enh, Delplanque:2024xst} for some low-spin examples. Massive (bi)-gravity \cite{deRham:2010kj,Hassan:2011zd,deRham:2014zqa} should also admit a chiral and the Proca-like description, see \cite{Alexandrov:2012yv} for the steps in the latter direction, which originated from the chiral formulation of (self-dual) gravity, see \emph{e.g.} \cite{Krasnov:2010olp,Krasnov:2016emc}.

As discussed above, parity is not easily controlled with type-$(m,n)$ fields, except in the symmetric approach where $m = n$. Likewise, imposing a reality condition on the Lagrangian is straightforward when using a real tensor field in the symmetric approach, but this is not the case in general. Therefore, save for the $m=n$ case, our Lagrangians describe a doubled number of degrees of freedom, which may still not be an obstacle for applications. One could try to achieve parity and reality by employing two fields of type $(m,n)$ and $(n,m)$ in a parity-even combination. For example, it is proposed in \cite{Curtright_1980413} to describe a massive spin-$s$ particle via a tensor $T_{[\mu\nu]\nu(s-1)}$, which takes values in the representation $(s+1,s-1)\oplus (s-1,s+1)$. However, much less is known about interactions within this manifestly parity-invariant approach. On the contrary, within the chiral approach \cite{Ochirov:2022nqz}, \emph{i.e.} $m=2s$, $n=0$, the simplest, minimal, interactions lead to infinitely many correct amplitudes immediately. Other amplitudes can be corrected order by order, see \emph{e.g.} \cite{Cangemi:2023bpe,Cangemi:2023ysz}. Therefore, the reality/parity conditions for the Lagrangian in any non-symmetric approach can be imposed order by order, which is not an obstacle for applications, \emph{e.g.} to the problem of black hole scattering.

There is a number of extensions of the present paper that one can envisage. Firstly, there should exist an extension of the Zinoviev approach \cite{Zinoviev:2001dt} to all $(m,n)$-descriptions, see also \cite{Fegebank:2024yft}. The field content should consists of double-traceless fields $\Phi_s$, ... $\Phi_\lambda$, where $\Phi_s$ is of type $(m,n)$, $m+n=2s$, but contains a trace, which is of type $(m-2,n-2)$. The last field $\Phi_\lambda$ is of type $(m-n,0)$ and is traceless. The action is fixed by requiring it be gauge-invariant under the most general gauge transformations with irreducible parameters $\xi_{s-1}$, ... $\!\!$, $\xi_{\lambda}$, where $\xi_{s-1}$ is of type $(m-1,n-1)$. In this gauge invariant formulation it should be possible to let $s$ be any real number, which should give a description of the continuous-spin particles, which is similar to \cite{Metsaev:2016lhs,Khabarov:2017lth}, see \emph{e.g.} \cite{Bekaert:2017khg} for a review. Secondly, a more natural realization of the double-trace constraint is the frame-like approach to massive fields \cite{Zinoviev:2008ze,Khabarov:2017lth,Khabarov:2019dvi}, which should also admit an extension to the $(m,n)$-descriptions, all of these fields are already present in the (non-minimal) unfolded formulation \cite{Ponomarev:2010st}. Thirdly, the Lagrangians for the $(m,n)$-description should admit an extension to $(A)dS_4$ spacetimes. Lastly, while in the paper we have used the special features of the $d=4$ world, an extension to higher dimensions may also be possible as the recent example \cite{Basile:2024raj} shows for massless fields.

%%%%%%%%%%%%%%%%%%%%%%%%%%%%%%%%%%%%%%%%%%%%%%%%%%%%%%%%%%%%%
\section*{Acknowledgments}
\label{sec:Aknowledgements}
%%%%%%%%%%%%%%%%%%%%%%%%%%%%%%%%%%%%%%%%%%%%%%%%%%%%%%%%%%%%%
The work of the author was supported by UMONS stipend ``Bourse d'encouragement doctorale FRIA/\-FRESH''. The work of the author was partially supported by the European Research Council (ERC) under the European Union’s Horizon 2020 research and innovation programme (grant agreement No 101002551). The author wishes to express deep gratitude to Evgeny Skvortsov for conceiving the idea for this project, for the many insightful discussions that contributed to its completion, and for his valuable guidance in the preparation of this manuscript.
%%%%%%%%%%%%%%%%%%%%%%%%%%%%%%%%%%%%%%%%%%%%%%%%%%%%%%%%%%%%%

\appendix

\section{Details of the proof}
\label{sec:app:proof}

Let us present the proof of the result \eqref{eq:solution}. The values of the coefficients of the action are fixed by the analysis of the consequences \eqref{eq:constraint_generic_expr} for $j$ from $\lambda$ to $s-2$ and in order to impose $\divg\eom_s = 0 \Leftrightarrow \divg\Phi_s = 0$. The procedure has been already explained in Section \ref{sec:Action}. Let us only point out that we will have to analyze the matrices $\mathbb{A}_j$ from $j=s-1$ to $j=\lambda$, which implies that the only role of $\lambda$ is to know where we need to stop. 

As it is easy to see from the matrices presented as examples in Section \ref{sec:Action}, from $\mathbb{A}_j$ we determine $b_{j+1}$, $a_j$ and $g_j$, with the help of the results for higher ranks, which has been found before. It leads naturally to a recursive proof of the formulas \eqref{eq:solution}: we assume that they are true for $r\geq j+1$ ($j+2$ for $b_r$) and we use the rank-$j$ constraint to show that they are then true for $r=j$ ($j+1$ for $b_r$). Note that the possibility to solve in this way shows that the coefficients in the action do not depend on $\lambda$, only the number of fields depends on it.

The idea now is to establish the recursion based on the structure of the row reduction of $\mathbb{A}_j$, \emph{i.e.} how the coefficients $b_{j+1}$, $a_j$ and $g_j$ depend on the previous ones $b_{r+1}$, $a_{r}$ and $g_{r}$ with $r\geq j+1$. Let us rewrite the matrix \eqref{eq:matrix_generic} in a suggestive form
\begin{align}
\left(\begin{array}{cc|ccc|ccc|ccc|ccc|ccc|ccc|ccc}
\alpha & \gamma & \beta & 0 & 0 &  &  &  &  &  &  &  &  &  &  &  &  &  &  &  \\ \hline
1 & 0 & \alpha & \gamma & 0 & \varepsilon & 0 & 0 &  &  &  &  &  &  &  &  &  &  &  &  \\
0 & 1 & 0 & \alpha & \gamma & 0 & \varepsilon & 0 &  &  &  &  &  &  &  &  &  &  &  &  \\ \hline
 &  & 0 & \!\!\!-m\!\!\! & 0 & \alpha & \gamma & 0 & \varepsilon & 0 & 0 &  &  &  &  &  &  &  &  &  \\
 &  & 0 & 0 & \!\!\!-m\!\!\! & 0 & \alpha & \gamma & 0 & \varepsilon & 0 &  &  &  &  &  &  &  &  &  \\ \hline
 &  &  &  &  & 0 & \!\!\!-m\!\!\! & 0 & \alpha & \gamma & 0 & \varepsilon & 0 & 0 &  &  &  &  &  &  \\
 &  &  &  &  & 0 & 0 & \!\!\!-m\!\!\! & 0 & \alpha & \gamma & 0 & \varepsilon & 0 &  &  &  &  &  &  \\ \hline
 &  &  &  &  &  &  &  &  0 & \!\!\!-m\!\!\! & 0 & \alpha & \gamma & 0 & \varepsilon & 0 & 0  &  &  &  &  &  & \\
 &  &  &  &  &  &  &  & 0 & 0 & \!\!\!-m\!\!\! & 0 & \alpha & \gamma & 0 & \varepsilon & 0  &  &  &  &  &  & \\ \hline
 &  &  &  &  &  &  &  &  &  &  &  0 & \!\!\!-m\!\!\! & 0 & \alpha & \gamma & 0 & \varepsilon & 0 & 0  &  &  & \\
 &  &  &  &  &  &  &  &  &  &  & 0 & 0 & \!\!\!-m\!\!\! & 0 & \alpha & \gamma & 0 & \varepsilon & 0 &  &  & \\ \hline
 &  &  &  &  &  &  &  &  &  &  &  &  &  &  0 & \!\!\!-m\!\!\! & 0 & \alpha' & \gamma & 0 & \varepsilon & 0 & 0 \\
 &  &  &  &  &  &  &  &  &  &  &  &  &  & 0 & 0 & \!\!\!-m\!\!\! & 0 & \alpha' & \gamma & 0 & \varepsilon & 0 \\ \hline
 &  &  &  &  &  &  &  &  &  &  &  &  &  & & & & 0 & \!\!\!-m\!\!\! & 0 & \alpha' & \gamma' & 0 \\
 &  &  &  &  &  &  &  &  &  &  &  &  &  & & & & 0 & 0 & \!\!\!-m\!\!\! & 0 & \alpha' & \gamma'
\end{array}\right) \, .
\label{eq:matrix_generic_proof}
\end{align}
Here, we replaced $\eta$ by its value (one) and $\zeta$'s by their value ($-m$); primed symbols contain coefficients that have not been fixed at the previous stages. We assume as known the values of the coefficients $b_{r+1}$, $a_{r}$ and $g_{r}$ with $r\geq j+1$. The recursive formulas giving the rank-$j$ coefficients in terms of the rank-$(j+1)$ ones will be obtained by requiring a linear combination of the lines of the matrix to give a line of zeros except for the last element. Let us focus on each column. Let us remind that the weights of each line of the matrix in the linear combination are $A^j_{\ell,k}$. We normalize the linear combination by choosing $A^j_{0,0} = 1$. We begin with the last column, which needs to give a non-zero number. This column contains zeros everywhere except $\gamma^j_0$ at the end. Therefore, $\gamma^j_0$ has to be non-zero, which means that $g_j$ has to be non-zero too, for any $j$. Note that this result is not surprising because $g_j$ are the coefficients of the mass terms. Next, we have to impose that the linear combination gives zero for each remaining column.

The last column of each block-column (except for the last one) contains only two non-zero elements: one in the last line of a block-line, $\gamma^j_{\ell}$, and one in the last line of the next block-line, $-m$. Inside $\gamma^j_{\ell}$, there is $g_{j+\ell}$, whose value is known from the previous matrix because the smallest $\ell$ considered here is $\ell = 1$. The linear combination for these columns gives
\begin{align}
A^j_{\ell,0}m^2g_{j+\ell} - A^j_{\ell-1,0}m = 0 \, , \quad \forall \ell\in \bbl 1, s-j-2 \bbr \, .
\end{align}
Because we fixed $A^j_{0,0} = 1$, this relation fixes $A^j_{1,0}$, which will then fix $A^j_{2,0}$, \emph{etc}. By recursion, we find
\begin{align}
A^j_{\ell,0} = \prod_{i=1}^{\ell}\big(mg_{j+i}\big)^{-1} \, , \quad \forall \ell\in \bbl 1, s-j-2 \bbr \, .
\label{eq:Ajl0_formula}
\end{align}
Then, for $\ell=s-j$ we have
\begin{align}
A^j_{s-j,0}m^2g_{s} + A^j_{s-j-2,0} = 0 \, .
\end{align}
By using \eqref{eq:Ajl0_formula} and the normalization $g_s = -1$, we obtain
\begin{align}
A^j_{s-j,0} = m^{-2}\prod_{i=1}^{s-j-2}\big(mg_{j+i}\big)^{-1} \, .
\label{eq:Aj(s-j)0_formula}
\end{align}
We started from bottom to top to find all the weights for $k=0$. Let us find the weights for $k=1$ from top to bottom now. The linear combination for first column of the first block-column says
\begin{align}
&A^j_{s-j,0}\alpha^j_{s-j} + A^j_{s-j-2,1} = 0 \label{eq:Aj(s-j-2)1_rec} \\
\Leftrightarrow \quad &A^j_{s-j-2,1} = m^{-2}\tfrac{(s+j)(s-j-1)}{2s}\prod_{i=1}^{s-j-2}\big(mg_{j+i}\big)^{-1} \, ,
\label{eq:Aj(s-j-2)1_formula}
\end{align}
where we used \eqref{eq:normalization}, \eqref{eq:def_greek}, \eqref{eq:b_s} and \eqref{eq:Aj(s-j)0_formula}. The first column of the $\ell^{\text{th}}$ block-column is
\begin{align}
&A^j_{\ell+1,1}\varepsilon^j_{\ell+1} + A^j_{\ell,1}\alpha^j_{\ell} = 0 \, , \quad \forall \ell \in \bbl 0, s-j-3 \bbr \\
\Leftrightarrow \quad &A^j_{\ell,1} = \tfrac{-m(\ell+1)(2j+\ell+2)}{2a_{j+\ell}+\ell(2j+\ell+1)b_{j+\ell}}A^j_{\ell+1,1} \, ,
\end{align}
where we used \eqref{eq:def_greek} to write the second line. Applying this formula recursively gives
\begin{align}
&A^j_{\ell,1} = A^j_{s-j-2,1}\prod_{i=\ell}^{s-j-3}\tfrac{-m(i+1)(2j+i+2)}{2a_{j+i}+i(2j+i+1)b_{j+i}} \\
\Leftrightarrow \quad &A^j_{\ell,1} = m^{-2}\tfrac{(s+j)(s-j-1)}{2s}\bigg(\prod_{i=1}^{\ell}(mg_{j+i})^{-1}\bigg)\prod_{i=1}^{s-j-3}\tfrac{-m(i+1)(2j+i+2)}{g_{j+i+1}\big(2a_{j+i}+i(2j+i+1)b_{j+i}\big)} \, ,\label{Aeleven}
\end{align}
where we used \eqref{eq:Aj(s-j-2)1_formula} to write the second line. Inserting the formulas \eqref{eq:solution_b_r_b_r+1}, \eqref{eq:solution_a_r} and \eqref{eq:solution_g_r} into \eqref{Aeleven} gives
\begin{align}
A^j_{\ell,1} = m^{-2}\tfrac{(\ell+1)(2j+\ell+2)}{(s-j-\ell-1)(s+j+\ell+2)}\prod_{i=1}^{\ell}(mg_{j+i})^{-1} \, , \quad \forall \ell \in \bbl 0, s-j-3 \bbr \, .
\label{eq:Ajl1_formula}
\end{align}
Note that we cannot use this formula for $\ell = 1$ and $\ell = 0$ because it relies on those coefficients in \eqref{eq:solution_b_r_b_r+1} and \eqref{eq:solution_a_r} that we have not yet obtained within the recursion. Next, the first column of the second last block-column implies
\begin{align}
&A^j_{2,1}\varepsilon^j_2 + A^j_{1,1}\alpha^j_1 = 0 \\
\Leftrightarrow \quad &\frac{A^j_{1,1}}{A^j_{2,1}} = -m\frac{2j+3}{a_{j+1}+(j+1)b_{j+1}} \, ,
\label{eq:Aj11/Aj21}
\end{align}
where we used \eqref{eq:def_greek}. Equivalently, the first column of the last block-column gives
\begin{align}
&A^j_{1,1}\varepsilon^j_1 + A^j_{0,1}\alpha^j_0 = 0 \\
\Leftrightarrow \quad &\frac{A^j_{0,1}}{A^j_{1,1}} = -m\frac{j+1}{a_{j}} \, .
\label{eq:Aj01/Aj11}
\end{align}
The second column of the third last block-column gives
\begin{align}
A^j_{3,0}\varepsilon^j_3 + A^j_{2,1}\gamma^j_2 + A^j_{2,0}\alpha^j_2 + A^j_{1,1}\zeta^j_1 = 0 \, .
\end{align}
Dividing it by $A^j_{2,1}$ and using \eqref{eq:def_greek}, \eqref{eq:Ajl1_formula}, \eqref{eq:Aj11/Aj21} and the formulas \eqref{eq:solution_b_r_b_r+1} (for $r\geq j+2$) and \eqref{eq:solution_a_r}, \eqref{eq:solution_g_r} (for $r\geq j+1$), gives
\begin{align}
b_{j+1} = \tfrac{2(j+3)}{(j+4)(s-j-3)(s+j+4)}\frac{1}{b_{j+2}} \, .
\end{align}
This proves the formula \eqref{eq:solution_b_r_b_r+1} for $r=j+1$, and, recursively, for all $r$. Note that we can use \eqref{eq:Ajl1_formula} for $\ell = 1$ too now. The second column of the second last block-column gives
\begin{align}
A^j_{2,0}\varepsilon^j_2 + A^j_{1,1}\gamma^j_1 + A^j_{1,0}\alpha^j_1 + A^j_{0,1}\zeta^j_0 = 0 \, .
\end{align}
Dividing it by $A^j_{1,1}$ and using \eqref{eq:def_greek}, \eqref{eq:Ajl1_formula}, \eqref{eq:Aj01/Aj11} and the formulas \eqref{eq:solution_b_r_b_r+1} (for all $r$ because the formula is proved now) and \eqref{eq:solution_a_r}, \eqref{eq:solution_g_r} (for $r\geq j+1$), gives
\begin{align}
a_j = (2j+3)b_j \, ,
\end{align}
which proves the formula \eqref{eq:solution_a_r} for $r=j$ and, hence, for all $r$. Note that we can use \eqref{eq:Ajl1_formula} for all values of $\ell$ now. Finally, the second column of the last block-column is
\begin{align}
A^j_{1,0}\varepsilon^j_1 + A^j_{0,1}\gamma^j_0 + A^j_{0,0}\alpha^j_0 = 0 \, .
\end{align}
By using \eqref{eq:def_greek}, \eqref{eq:Ajl1_formula}, the formula \eqref{eq:solution_g_r} for $r\geq j+1$ and the formulas \eqref{eq:solution_b_r_b_r+1} and \eqref{eq:solution_a_r} for all $r$ we obtain
\begin{align}
g_j = -\tfrac{(j+2)(s-j)(s+j+1)}{2(j+1)}b_j \, ,
\end{align}
which proves the formula \eqref{eq:solution_g_r}. Thus, we have obtained a system of recursive relations among all coefficients in the action and, therefore, they are all determined. 

Nevertheless, it is useful to get $b_j$ in a closed form, \emph{i.e.} to prove the formula \eqref{eq:solution_b_r}, which we do recursively again. Applying twice \eqref{eq:solution_b_r_b_r+1} gives
\begin{align}
b_r = \tfrac{(r+2)(r+4)(s-r-3)(s+r+4)}{(r+3)^2(s-r-2)(s+r+3)}b_{r+2} \, , \quad \forall r \in \bbl \lambda, s-j-4 \bbr \, .
\label{eq:b_r_b_r+2}
\end{align}
Simple iterations of \eqref{eq:b_r_b_r+2} yield  \eqref{eq:solution_b_r}.

We extracted enough information from the matrix to find all the coefficients. Nevertheless, we still need to check all other linear combinations. The first column of the second block-column gives
\begin{align}
A^j_{s-j,0}\beta^j_{s-j} + A^j_{s-j-2,1}\alpha^j_{s-j-2} = 0 \, .
\end{align}
By inserting \eqref{eq:Aj(s-j-2)1_rec} we find
\begin{align}
\beta^j_{s-j} = \alpha^j_{s-j}\alpha^j_{s-j-2} \, ,
\end{align}
which is identically satisfied by using \eqref{eq:def_greek} with \eqref{eq:normalization} and \eqref{eq:solution_s-2}. Then, the second column of the second block-column implies
\begin{align}
A^j_{s-j-2,1}\gamma^j_{s-j-2} + A^j_{s-j-2,0}\alpha^j_{s-j-2} + A^j_{s-j-3,1}\zeta^j_{s-j-3} = 0 \, .
\end{align}
By using \eqref{eq:def_greek}, \eqref{eq:solution_b_r_b_r+1}, \eqref{eq:solution_a_r}, \eqref{eq:Ajl0_formula}, \eqref{eq:Aj(s-j-2)1_formula} and \eqref{eq:Ajl1_formula} one can show that this relation is identically satisfied. Finally, the second column of the $\ell^\text{th}$ block-column implies
\begin{align}
A^j_{\ell+2,0}\varepsilon^j_{\ell+2} + A^j_{\ell+1,1}\gamma^j_{\ell+1} + A^j_{\ell+1,0}\alpha^j_{\ell+1} + A^j_{\ell,1}\zeta^j_{\ell} = 0 \, .
\end{align}
Inserting \eqref{eq:def_greek}, \eqref{eq:solution_a_r}, \eqref{eq:solution_g_r}, \eqref{eq:Ajl0_formula} and \eqref{eq:Ajl1_formula} one can show that it is identically satisfied.

%%%%%%%%%%%%%%%%%%%%%%%%%%%%%%%%%%%%%%%%%%%%%%%%%%%%%%%%%%%%%

\newpage
\footnotesize
\providecommand{\href}[2]{#2}\begingroup\raggedright\endgroup


\begin{thebibliography}{10}

\bibitem{Buonanno:2022pgc}
A.~Buonanno, M.~Khalil, D.~O'Connell, R.~Roiban, M.~P. Solon, and M.~Zeng, ``{Snowmass White Paper: Gravitational Waves and Scattering Amplitudes},'' in {\em {Snowmass 2021}}.
\newblock 4, 2022.
\newblock \href{http://arxiv.org/abs/2204.05194}{{\ttfamily arXiv:2204.05194 [hep-th]}}.

\bibitem{Guevara:2018wpp}
A.~Guevara, A.~Ochirov, and J.~Vines, ``{Scattering of Spinning Black Holes from Exponentiated Soft Factors},'' \href{http://dx.doi.org/10.1007/JHEP09(2019)056}{{\em JHEP} {\bfseries 09} (2019) 056},
\href{http://arxiv.org/abs/1812.06895}{{\ttfamily arXiv:1812.06895 [hep-th]}}.
%%CITATION = ARXIV:1812.06895;%%.

\bibitem{Chung:2018kqs}
M.-Z. Chung, Y.-T. Huang, J.-W. Kim, and S.~Lee, ``{The simplest massive S-matrix: from minimal coupling to Black Holes},'' \href{http://dx.doi.org/10.1007/JHEP04(2019)156}{{\em JHEP} {\bfseries 04} (2019) 156},
\href{http://arxiv.org/abs/1812.08752}{{\ttfamily arXiv:1812.08752 [hep-th]}}.
%%CITATION = ARXIV:1812.08752;%%.

\bibitem{Cangemi:2022bew}
L.~Cangemi, M.~Chiodaroli, H.~Johansson, A.~Ochirov, P.~Pichini, and E.~Skvortsov, ``{Kerr Black Holes From Massive Higher-Spin Gauge Symmetry},'' \href{http://dx.doi.org/10.1103/PhysRevLett.131.221401}{{\em Phys. Rev. Lett.} {\bfseries 131} no.~22, (2023) 221401}, \href{http://arxiv.org/abs/2212.06120}{{\ttfamily arXiv:2212.06120 [hep-th]}}.

\bibitem{Cangemi:2023bpe}
L.~Cangemi, M.~Chiodaroli, H.~Johansson, A.~Ochirov, P.~Pichini, and E.~Skvortsov, ``{Compton Amplitude for Rotating Black Hole from QFT},'' \href{http://dx.doi.org/10.1103/PhysRevLett.133.071601}{{\em Phys. Rev. Lett.} {\bfseries 133} no.~7, (2024) 071601}, \href{http://arxiv.org/abs/2312.14913}{{\ttfamily arXiv:2312.14913 [hep-th]}}.

\bibitem{Skvortsov:2023jbn}
E.~Skvortsov and M.~Tsulaia, ``{Cubic action for Spinning Black Holes from massive higher-spin gauge symmetry},'' \href{http://arxiv.org/abs/2312.08184}{{\ttfamily arXiv:2312.08184 [hep-th]}}.

\bibitem{Cangemi:2023ysz}
L.~Cangemi, M.~Chiodaroli, H.~Johansson, A.~Ochirov, P.~Pichini, and E.~Skvortsov, ``{From higher-spin gauge interactions to Compton amplitudes for root-Kerr},'' \href{http://dx.doi.org/10.1007/JHEP09(2024)196}{{\em JHEP} {\bfseries 09} (2024) 196}, \href{http://arxiv.org/abs/2311.14668}{{\ttfamily arXiv:2311.14668 [hep-th]}}.

\bibitem{Wigner:1939cj}
E.~P. Wigner, ``{On Unitary Representations of the Inhomogeneous Lorentz Group},'' \href{http://dx.doi.org/10.2307/1968551}{{\em Annals Math.} {\bfseries 40} (1939) 149--204}.
[Reprint: Nucl. Phys. Proc. Suppl.6,9(1989)].
%%CITATION = ANMAA,40,149;%%.

\bibitem{Bekaert:2006py}
X.~Bekaert and N.~Boulanger, ``The unitary representations of the poincare group in any spacetime dimension,''
\href{http://arxiv.org/abs/hep-th/0611263}{{\ttfamily hep-th/0611263}}.
%%CITATION = HEP-TH/0611263;%%.

\bibitem{Basile:2016aen}
T.~Basile, X.~Bekaert, and N.~Boulanger, ``{Mixed-symmetry fields in de Sitter space: a group theoretical glance},'' \href{http://dx.doi.org/10.1007/JHEP05(2017)081}{{\em JHEP} {\bfseries 05} (2017) 081}, \href{http://arxiv.org/abs/1612.08166}{{\ttfamily arXiv:1612.08166 [hep-th]}}.

\bibitem{Bekaert:2017khg}
X.~Bekaert and E.~D. Skvortsov, ``{Elementary particles with continuous spin},'' \href{http://dx.doi.org/10.1142/S0217751X17300198}{{\em Int. J. Mod. Phys. A} {\bfseries 32} no.~23n24, (2017) 1730019}, \href{http://arxiv.org/abs/1708.01030}{{\ttfamily arXiv:1708.01030 [hep-th]}}.

\bibitem{Metsaev:1991mt}
R.~R. Metsaev, ``{Poincare invariant dynamics of massless higher spins: Fourth order analysis on mass shell},''
{\em Mod. Phys. Lett.} {\bfseries A6} (1991) 359--367.
%%CITATION = MPLAE,A6,359;%%.

\bibitem{Metsaev:1991nb}
R.~R. Metsaev, ``{$S$ matrix approach to massless higher spins theory. 2: The Case of internal symmetry},''
{\em Mod. Phys. Lett.} {\bfseries A6} (1991) 2411--2421.
%%CITATION = MPLAE,A6,2411;%%.

\bibitem{Ponomarev:2016lrm}
D.~Ponomarev and E.~D. Skvortsov, ``{Light-Front Higher-Spin Theories in Flat Space},'' {\em J. Phys.} {\bfseries A50} no.~9, (2017) 095401,
\href{http://arxiv.org/abs/1609.04655}{{\ttfamily arXiv:1609.04655 [hep-th]}}.
%%CITATION = ARXIV:1609.04655;%%.

\bibitem{Skvortsov:2018jea}
E.~D. Skvortsov, T.~Tran, and M.~Tsulaia, ``{Quantum Chiral Higher Spin Gravity},'' {\em Phys. Rev. Lett.} {\bfseries 121} no.~3, (2018) 031601,
\href{http://arxiv.org/abs/1805.00048}{{\ttfamily arXiv:1805.00048 [hep-th]}}.
%%CITATION = ARXIV:1805.00048;%%.

\bibitem{Skvortsov:2020wtf}
E.~Skvortsov, T.~Tran, and M.~Tsulaia, ``{More on Quantum Chiral Higher Spin Gravity},'' {\em Phys. Rev.} {\bfseries D101} no.~10, (2020) 106001,
\href{http://arxiv.org/abs/2002.08487}{{\ttfamily arXiv:2002.08487 [hep-th]}}.
%%CITATION = ARXIV:2002.08487;%%.

\bibitem{Ponomarev:2017nrr}
D.~Ponomarev, ``{Chiral Higher Spin Theories and Self-Duality},'' {\em JHEP} {\bfseries 12} (2017) 141,
\href{http://arxiv.org/abs/1710.00270}{{\ttfamily arXiv:1710.00270 [hep-th]}}.
%%CITATION = ARXIV:1710.00270;%%.

\bibitem{Krasnov:2021nsq}
K.~Krasnov, E.~Skvortsov, and T.~Tran, ``{Actions for Self-dual Higher Spin Gravities},''
\href{http://arxiv.org/abs/2105.12782}{{\ttfamily arXiv:2105.12782 [hep-th]}}.
%%CITATION = ARXIV:2105.12782;%%.

\bibitem{Sharapov:2022wpz}
A.~Sharapov, E.~Skvortsov, and R.~Van~Dongen, ``{Chiral Higher Spin Gravity and Convex Geometry},'' \href{http://arxiv.org/abs/2209.01796}{{\ttfamily arXiv:2209.01796 [hep-th]}}.

\bibitem{Bekaert:2022poo}
X.~Bekaert, N.~Boulanger, A.~Campoleoni, M.~Chiodaroli, D.~Francia, M.~Grigoriev, E.~Sezgin, and E.~Skvortsov, ``{Snowmass White Paper: Higher Spin Gravity and Higher Spin Symmetry},'' \href{http://arxiv.org/abs/2205.01567}{{\ttfamily arXiv:2205.01567 [hep-th]}}.

\bibitem{Proca_1936}
A.~Proca, ``Sur la th\'eorie ondulatoire des \'electrons positifs et n\'egatifs,'' \href{http://dx.doi.org/10.1051/jphysrad:0193600708034700}{{\em J. Phys. Radium} {\bfseries 7} no.~8, (1936) 347--353}.

\bibitem{Fierz:1939ix}
M.~Fierz and W.~Pauli, ``{On relativistic wave equations for particles of arbitrary spin in an electromagnetic field},'' \href{http://dx.doi.org/10.1098/rspa.1939.0140}{{\em Proc. Roy. Soc. Lond. A} {\bfseries 173} (1939) 211--232}.

\bibitem{Singh:1974qz}
L.~P.~S. Singh and C.~R. Hagen, ``{Lagrangian formulation for arbitrary spin. 1. The boson case},''
\href{http://dx.doi.org/10.1103/PhysRevD.9.898}{{\em Phys. Rev.} {\bfseries D9} (1974) 898--909}.
%%CITATION = PHRVA,D9,898;%%.

\bibitem{Singh:1974rc}
L.~P.~S. Singh and C.~R. Hagen, ``{Lagrangian formulation for arbitrary spin. 2. The fermion case},''
\href{http://dx.doi.org/10.1103/PhysRevD.9.910}{{\em Phys. Rev.} {\bfseries D9} (1974) 910--920}.
%%CITATION = PHRVA,D9,910;%%.

\bibitem{Zinoviev:2001dt}
Y.~M. Zinoviev, ``{On massive high spin particles in AdS},'' \href{http://arxiv.org/abs/hep-th/0108192}{{\ttfamily arXiv:hep-th/0108192}}.

\bibitem{Zinovev:1983kh}
Y.~M. Zinovev, ``{GAUGE INVARIANT DESCRIPTION OF MASSIVE HIGH SPIN PARTICLES},'' {\em preprint IFVE-83-91} (6, 1983) .

\bibitem{Klishevich:1997pd}
S.~M. Klishevich and Y.~M. Zinovev, ``{On electromagnetic interaction of massive spin-2 particle},'' {\em Phys. Atom. Nucl.} {\bfseries 61} (1998) 1527--1537, \href{http://arxiv.org/abs/hep-th/9708150}{{\ttfamily arXiv:hep-th/9708150}}.

\bibitem{Pashnev:1989gm}
A.~I. Pashnev, ``{Composite Systems and Field Theory for a Free Regge Trajectory},'' \href{http://dx.doi.org/10.1007/BF01017664}{{\em Theor. Math. Phys.} {\bfseries 78} (1989) 272--277}.

\bibitem{Fegebank:2024yft}
J.~H. Fegebank and S.~M. Kuzenko, ``{On equivalence of gauge-invariant models for massive integer-spin fields},'' \href{http://arxiv.org/abs/2406.02573}{{\ttfamily arXiv:2406.02573 [hep-th]}}.

\bibitem{Buchbinder:2005ua}
I.~L. Buchbinder and V.~A. Krykhtin, ``{Gauge invariant Lagrangian construction for massive bosonic higher spin fields in D dimensions},'' \href{http://dx.doi.org/10.1016/j.nuclphysb.2005.07.035}{{\em Nucl. Phys. B} {\bfseries 727} (2005) 537--563}, \href{http://arxiv.org/abs/hep-th/0505092}{{\ttfamily arXiv:hep-th/0505092}}.

\bibitem{Bekaert:2003uc}
X.~Bekaert, I.~L. Buchbinder, A.~Pashnev, and M.~Tsulaia, ``{On higher spin theory: Strings, BRST, dimensional reductions},'' \href{http://dx.doi.org/10.1088/0264-9381/21/10/018}{{\em Class. Quant. Grav.} {\bfseries 21} (2004) S1457--1464}, \href{http://arxiv.org/abs/hep-th/0312252}{{\ttfamily arXiv:hep-th/0312252}}.

\bibitem{Francia:2010qp}
D.~Francia, ``{String theory triplets and higher-spin curvatures},'' {\em Phys.Lett.} {\bfseries B690} (2010) 90--95,
\href{http://arxiv.org/abs/1001.5003}{{\ttfamily arXiv:1001.5003 [hep-th]}}.
%%CITATION = ARXIV:1001.5003;%%.

\bibitem{Buchbinder:2007ix}
I.~L. Buchbinder, V.~A. Krykhtin, and H.~Takata, ``{Gauge invariant Lagrangian construction for massive bosonic mixed symmetry higher spin fields},'' \href{http://dx.doi.org/10.1016/j.physletb.2007.09.033}{{\em Phys. Lett. B} {\bfseries 656} (2007) 253--264}, \href{http://arxiv.org/abs/0707.2181}{{\ttfamily arXiv:0707.2181 [hep-th]}}.

\bibitem{Buchbinder:2008ss}
I.~L. Buchbinder and A.~V. Galajinsky, ``{Quartet unconstrained formulation for massive higher spin fields},'' \href{http://dx.doi.org/10.1088/1126-6708/2008/11/081}{{\em JHEP} {\bfseries 11} (2008) 081}, \href{http://arxiv.org/abs/0810.2852}{{\ttfamily arXiv:0810.2852 [hep-th]}}.

\bibitem{Kaparulin:2012px}
D.~S. Kaparulin, S.~L. Lyakhovich, and A.~A. Sharapov, ``{Consistent interactions and involution},'' \href{http://dx.doi.org/10.1007/JHEP01(2013)097}{{\em JHEP} {\bfseries 01} (2013) 097}, \href{http://arxiv.org/abs/1210.6821}{{\ttfamily arXiv:1210.6821 [hep-th]}}.

\bibitem{Kazinski:2005eb}
P.~O. Kazinski, S.~L. Lyakhovich, and A.~A. Sharapov, ``{Lagrange structure and quantization},'' \href{http://dx.doi.org/10.1088/1126-6708/2005/07/076}{{\em JHEP} {\bfseries 07} (2005) 076}, \href{http://arxiv.org/abs/hep-th/0506093}{{\ttfamily arXiv:hep-th/0506093}}.

\bibitem{Metsaev:2012uy}
R.~R. Metsaev, ``{BRST-BV approach to cubic interaction vertices for massive and massless higher-spin fields},'' \href{http://dx.doi.org/10.1016/j.physletb.2013.02.009}{{\em Phys. Lett. B} {\bfseries 720} (2013) 237--243}, \href{http://arxiv.org/abs/1205.3131}{{\ttfamily arXiv:1205.3131 [hep-th]}}.

\bibitem{Francia:2007ee}
D.~Francia, ``{Geometric Lagrangians for massive higher-spin fields},'' \href{http://dx.doi.org/10.1016/j.nuclphysb.2007.12.002}{{\em Nucl. Phys. B} {\bfseries 796} (2008) 77--122}, \href{http://arxiv.org/abs/0710.5378}{{\ttfamily arXiv:0710.5378 [hep-th]}}.

\bibitem{Francia:2008ac}
D.~Francia, ``{Geometric massive higher spins and current exchanges},'' \href{http://dx.doi.org/10.1002/prop.200810547}{{\em Fortsch. Phys.} {\bfseries 56} (2008) 800--808}, \href{http://arxiv.org/abs/0804.2857}{{\ttfamily arXiv:0804.2857 [hep-th]}}.

\bibitem{Curtright_1980413}
T.~L. Curtright and P.~G. Freund, ``Massive dual fields,'' \href{http://dx.doi.org/https://doi.org/10.1016/0550-3213(80)90174-1}{{\em Nuclear Physics B} {\bfseries 172} (1980) 413--424}.

\bibitem{González_2008}
B.~González, A.~Khoudeir, R.~Montemayor, and L.~Urrutia, ``Duality for massive spin two theories in arbitrary dimensions,'' \href{http://dx.doi.org/10.1088/1126-6708/2008/09/058}{{\em JHEP} {\bfseries 09} (2008) 058}, \href{http://arxiv.org/abs/0806.3200v2}{{\ttfamily arXiv:0806.3200v2 [hep-th]}}.

\bibitem{Hell_2022}
A.~Hell, ``On the duality of massive kalb-ramond and proca fields,'' \href{http://dx.doi.org/10.1088/1475-7516/2022/01/056}{{\em Journal of Cosmology and Astroparticle Physics} {\bfseries 01} (2022) 056}, \href{http://arxiv.org/abs/2109.05030v2}{{\ttfamily arXiv:2109.05030v2 [hep-th]}}.

\bibitem{Abakumova:2023wve}
V.~A. Abakumova and S.~L. Lyakhovich, ``{Dualisation of free fields},'' \href{http://dx.doi.org/10.1016/j.aop.2023.169322}{{\em Annals Phys.} {\bfseries 453} (2023) 169322}, \href{http://arxiv.org/abs/2303.02616}{{\ttfamily arXiv:2303.02616 [hep-th]}}.

\bibitem{Metsaev:2005ar}
R.~R. Metsaev, ``{Cubic interaction vertices of massive and massless higher spin fields},'' \href{http://dx.doi.org/10.1016/j.nuclphysb.2006.10.002}{{\em Nucl. Phys. B} {\bfseries 759} (2006) 147--201}, \href{http://arxiv.org/abs/hep-th/0512342}{{\ttfamily arXiv:hep-th/0512342}}.

\bibitem{Metsaev:2007rn}
R.~R. Metsaev, ``{Cubic interaction vertices for fermionic and bosonic arbitrary spin fields},'' \href{http://dx.doi.org/10.1016/j.nuclphysb.2012.01.022}{{\em Nucl. Phys. B} {\bfseries 859} (2012) 13--69}, \href{http://arxiv.org/abs/0712.3526}{{\ttfamily arXiv:0712.3526 [hep-th]}}.

\bibitem{Metsaev:2022yvb}
R.~R. Metsaev, ``{Interacting massive and massless arbitrary spin fields in 4d flat space},'' \href{http://arxiv.org/abs/2206.13268}{{\ttfamily arXiv:2206.13268 [hep-th]}}.

\bibitem{Chalmers:1997ui}
G.~Chalmers and W.~Siegel, ``{Simplifying algebra in Feynman graphs, Part I: Spinors},'' \href{http://dx.doi.org/10.1103/PhysRevD.59.045012}{{\em Phys. Rev.} {\bfseries D59} (1999) 045012},
\href{http://arxiv.org/abs/hep-ph/9708251}{{\ttfamily arXiv:hep-ph/9708251 [hep-ph]}}.
%%CITATION = HEP-PH/9708251;%%.

\bibitem{Chalmers:2001cy}
G.~Chalmers and W.~Siegel, ``{Simplifying algebra in Feynman graphs. 3. Massive vectors},'' \href{http://dx.doi.org/10.1103/PhysRevD.63.125027}{{\em Phys. Rev.} {\bfseries D63} (2001) 125027},
\href{http://arxiv.org/abs/hep-th/0101025}{{\ttfamily arXiv:hep-th/0101025 [hep-th]}}.
%%CITATION = HEP-TH/0101025;%%.

\bibitem{Ochirov:2022nqz}
A.~Ochirov and E.~Skvortsov, ``{Chiral Approach to Massive Higher Spins},'' \href{http://dx.doi.org/10.1103/PhysRevLett.129.241601}{{\em Phys. Rev. Lett.} {\bfseries 129} no.~24, (2022) 241601}, \href{http://arxiv.org/abs/2207.14597}{{\ttfamily arXiv:2207.14597 [hep-th]}}.

\bibitem{Koutrolikos_2021}
K.~Koutrolikos, ``Superspace formulation of massive half-integer superspin,'' \href{http://dx.doi.org/10.1007/JHEP03(2021)254}{{\em JHEP} {\bfseries 03} (2021) 254}, \href{http://arxiv.org/abs/2012.12225v2}{{\ttfamily arXiv:2012.12225v2 [hep-th]}}.

\bibitem{Delplanque:2024enh}
W.~Delplanque and E.~Skvortsov, ``{Symmetric vs. chiral approaches to massive fields with spin},'' \href{http://dx.doi.org/10.1088/1361-6382/ad8f27}{{\em Class. Quant. Grav.} {\bfseries 41} (2024) 245018}, \href{http://arxiv.org/abs/2405.13706v2}{{\ttfamily arXiv:2405.13706v2 [hep-th]}}.

\bibitem{Zinoviev:2006im}
Y.~M. Zinoviev, ``{On massive spin 2 interactions},'' \href{http://dx.doi.org/10.1016/j.nuclphysb.2007.02.005}{{\em Nucl. Phys. B} {\bfseries 770} (2007) 83--106}, \href{http://arxiv.org/abs/hep-th/0609170}{{\ttfamily arXiv:hep-th/0609170}}.

\bibitem{Zinoviev:2008ck}
Y.~M. Zinoviev, ``{On spin 3 interacting with gravity},'' \href{http://dx.doi.org/10.1088/0264-9381/26/3/035022}{{\em Class. Quant. Grav.} {\bfseries 26} (2009) 035022}, \href{http://arxiv.org/abs/0805.2226}{{\ttfamily arXiv:0805.2226 [hep-th]}}.

\bibitem{Zinoviev:2009hu}
Y.~M. Zinoviev, ``{On massive spin 2 electromagnetic interactions},'' \href{http://dx.doi.org/10.1016/j.nuclphysb.2009.04.027}{{\em Nucl. Phys. B} {\bfseries 821} (2009) 431--451}, \href{http://arxiv.org/abs/0901.3462}{{\ttfamily arXiv:0901.3462 [hep-th]}}.

\bibitem{Zinoviev:2010cr}
Y.~M. Zinoviev, ``{Spin 3 cubic vertices in a frame-like formalism},'' \href{http://dx.doi.org/10.1007/JHEP08(2010)084}{{\em JHEP} {\bfseries 08} (2010) 084}, \href{http://arxiv.org/abs/1007.0158}{{\ttfamily arXiv:1007.0158 [hep-th]}}.

\bibitem{Buchbinder:2012iz}
I.~L. Buchbinder, T.~V. Snegirev, and Y.~M. Zinoviev, ``{Cubic interaction vertex of higher-spin fields with external electromagnetic field},'' \href{http://dx.doi.org/10.1016/j.nuclphysb.2012.07.012}{{\em Nucl. Phys. B} {\bfseries 864} (2012) 694--721}, \href{http://arxiv.org/abs/1204.2341}{{\ttfamily arXiv:1204.2341 [hep-th]}}.

\bibitem{Delplanque:2024xst}
W.~Delplanque and E.~Skvortsov, ``{Massive spin three-half field in a constant electromagnetic background},'' \href{http://dx.doi.org/10.1007/JHEP08(2024)173}{{\em JHEP} {\bfseries 08} (2024) 173}, \href{http://arxiv.org/abs/2406.14148}{{\ttfamily arXiv:2406.14148 [hep-th]}}.

\bibitem{deRham:2010kj}
C.~de~Rham, G.~Gabadadze, and A.~J. Tolley, ``{Resummation of Massive Gravity},'' \href{http://dx.doi.org/10.1103/PhysRevLett.106.231101}{{\em Phys. Rev. Lett.} {\bfseries 106} (2011) 231101}, \href{http://arxiv.org/abs/1011.1232}{{\ttfamily arXiv:1011.1232 [hep-th]}}.

\bibitem{Hassan:2011zd}
S.~F. Hassan and R.~A. Rosen, ``{Bimetric Gravity from Ghost-free Massive Gravity},'' \href{http://dx.doi.org/10.1007/JHEP02(2012)126}{{\em JHEP} {\bfseries 02} (2012) 126}, \href{http://arxiv.org/abs/1109.3515}{{\ttfamily arXiv:1109.3515 [hep-th]}}.

\bibitem{deRham:2014zqa}
C.~de~Rham, ``{Massive Gravity},'' \href{http://dx.doi.org/10.12942/lrr-2014-7}{{\em Living Rev. Rel.} {\bfseries 17} (2014) 7}, \href{http://arxiv.org/abs/1401.4173}{{\ttfamily arXiv:1401.4173 [hep-th]}}.

\bibitem{Alexandrov:2012yv}
S.~Alexandrov, K.~Krasnov, and S.~Speziale, ``{Chiral description of ghost-free massive gravity},'' \href{http://dx.doi.org/10.1007/JHEP06(2013)068}{{\em JHEP} {\bfseries 06} (2013) 068}, \href{http://arxiv.org/abs/1212.3614}{{\ttfamily arXiv:1212.3614 [hep-th]}}.

\bibitem{Krasnov:2010olp}
K.~Krasnov, ``{Plebanski Formulation of General Relativity: A Practical Introduction},'' \href{http://dx.doi.org/10.1007/s10714-010-1061-x}{{\em Gen. Rel. Grav.} {\bfseries 43} (2011) 1--15}, \href{http://arxiv.org/abs/0904.0423}{{\ttfamily arXiv:0904.0423 [gr-qc]}}.

\bibitem{Krasnov:2016emc}
K.~Krasnov, ``{Self-Dual Gravity},'' \href{http://dx.doi.org/10.1088/1361-6382/aa65e5}{{\em Class. Quant. Grav.} {\bfseries 34} no.~9, (2017) 095001}, \href{http://arxiv.org/abs/1610.01457}{{\ttfamily arXiv:1610.01457 [hep-th]}}.

\bibitem{Metsaev:2016lhs}
R.~R. Metsaev, ``{Continuous spin gauge field in (A)dS space},'' \href{http://dx.doi.org/10.1016/j.physletb.2017.02.027}{{\em Phys. Lett. B} {\bfseries 767} (2017) 458--464}, \href{http://arxiv.org/abs/1610.00657}{{\ttfamily arXiv:1610.00657 [hep-th]}}.

\bibitem{Khabarov:2017lth}
M.~V. Khabarov and Y.~M. Zinoviev, ``{Infinite (continuous) spin fields in the frame-like formalism},'' \href{http://dx.doi.org/10.1016/j.nuclphysb.2018.01.016}{{\em Nucl. Phys. B} {\bfseries 928} (2018) 182--216}, \href{http://arxiv.org/abs/1711.08223}{{\ttfamily arXiv:1711.08223 [hep-th]}}.

\bibitem{Zinoviev:2008ze}
Y.~M. Zinoviev, ``{Frame-like gauge invariant formulation for massive high spin particles},'' {\em Nucl. Phys.} {\bfseries B808} (2009) 185--204,
\href{http://arxiv.org/abs/0808.1778}{{\ttfamily arXiv:0808.1778 [hep-th]}}.
%%CITATION = 0808.1778;%%.

\bibitem{Khabarov:2019dvi}
M.~V. Khabarov and Y.~M. Zinoviev, ``{Massive higher spin fields in the frame-like multispinor formalism},'' \href{http://dx.doi.org/10.1016/j.nuclphysb.2019.114773}{{\em Nucl. Phys. B} {\bfseries 948} (2019) 114773}, \href{http://arxiv.org/abs/1906.03438}{{\ttfamily arXiv:1906.03438 [hep-th]}}.

\bibitem{Ponomarev:2010st}
D.~S. Ponomarev and M.~A. Vasiliev, ``{Frame-Like Action and Unfolded Formulation for Massive Higher-Spin Fields},''
\href{http://arxiv.org/abs/1001.0062}{{\ttfamily arXiv:1001.0062 [hep-th]}}.
%%CITATION = 1001.0062;%%.

\bibitem{Basile:2024raj}
T.~Basile, ``{Massless chiral fields in six dimensions},'' \href{http://arxiv.org/abs/2409.12800}{{\ttfamily arXiv:2409.12800 [hep-th]}}.

\end{thebibliography}
\end{document}